\documentclass [12pt,preprint]{aastex}
\usepackage{epsfig}
\usepackage{epstopdf}
\usepackage{natbib}
\usepackage{multirow}
\begin{document}

\newcommand{\xmm}{{\small \it XMM-Newton}}
\newcommand{\swift}{{\small \it Swift}}
\newcommand{\rosat}{{\small \it ROSAT}}
\newcommand{\gsim}{\hbox{\rlap{$^>$}$_\sim$}}
\newcommand{\lsim}{\hbox{\rlap{$^<$}$_\sim$}}
\newcommand \x {X-ray}
\newcommand{\xspec}{{\small \it Xspec}}
\newcommand{\pn}{{\small EPIC~PN}}

\title{X-ray Absorption of High Redshift Quasars}

\author{Assaf Eitan\altaffilmark{1}, Ehud Behar\altaffilmark{2}
}

\altaffiltext{1}{Physics Department, Technion, Haifa 32000, Israel.
sassafe@tx.technion.ac.il}
\altaffiltext{2}{Physics Department, Technion, Haifa 32000, Israel.
behar@physics.technion.ac.il}

\begin{abstract}

Soft X-ray photoelectric absorption of high-$z$ quasars has been known for two decades, but has no unambiguous astro-physical context. We construct the largest sample to date of 58 high redshift quasars ($z > 0.45$) selected from the \textit{XMM-Newton} archive based on a high photon count criterion ($> 1800$). We measure the optical depth $\tau$ at 0.5~keV, and find that 43\% of the quasars show significant absorption. We aim to find which physical parameters of the quasars drive their observed absorption, e.g., redshift, radio luminosity, radio loudness, or the X-ray luminosity. We compare the absorption behavior with redshift with the pattern expected if the diffuse intergalactic medium (IGM) is responsible for the observed absorption, and with a comparison sample of gamma ray burst (GRB) X-ray afterglows. 
Although the $z > 2$ quasar opacity is consistent with diffuse IGM absorption, many intermediate $z$ ($0.45 < z < 2$) quasars are not sufficiently absorbed for this scenario, and are appreciably less absorbed than GRBs. 
Only 10/37 quasars at $z < 2$ are absorbed, and only 5/30 radio quiet quasars are absorbed. 
We find a weak correlation between $\tau$ and $z$, and even a weaker correlation between $\tau$ and radio luminosity, which leads to the conclusion that although a diffuse IGM origin for the quasar absorption is unlikely, optical depth does seem to increase with redshift, roughly as $(1+z)^{2.2\pm0.6}$, tending at high-$z$ to $\tau \approx 0.4$, similar to the high-$z$ GRB values. 
This result can be explained by an ionized and clumpy IGM at $z < 2$, and a cold, diffuse IGM at higher redshift.
If, conversely, ascribed to local absorption at the quasar, and owing to the steep $L_x \propto (1+z)^{7.1\pm0.5}$ correlation in the present sample, the host column density scales as $N\mathrm{_{H}} \propto L_x^{0.7\pm0.1}$.

\end{abstract}


\maketitle

\section{Introduction} 
\label{into}



X-ray absorption of high redshift quasars has been widely studied in the last two decades. \cite{Wilkes_1992} were the first to report detection of excess absorption toward a high-$z$ quasar.  They used a \textit{ROSAT} spectrum of the radio loud quasar PKS\,0483-436 at $z = 2.85$.
More high-$z$ quasars were observed in the following years, and soft X-ray absorption was found in a growing number of objects. See \citet{Elvis_1994} for a compilation of observations from several observatories. These findings were surprising at the time, reversing the observed trend at lower redshift for X-ray absorption to decrease with luminosity and suggesting that perhaps damped Ly$\alpha$ systems along the line of sight, or intracluster material around the quasars is absorbing the X-rays \citep{Elvis_1994} . Thus, an intriguing question emerging from such results, is the identity of the absorber. Namely, is the absorption due to intervening systems along the line of sight to the source, or is it intrinsic to the source, and associated with physical processes in the quasar?
\cite{Fiore_1998}, using \textit{ROSAT} spectra, found that soft X-ray absorption is significantly more common in radio loud quasars (RLQs) than in radio quiet quasars (RQQs), 
implying perhaps that the RLQ jet is absorbing X-rays, although a viable physical mechanism that places atomic matter in the jet, but downstream from the X-ray source, is yet to be identified.
On the other hand, \citet{Fiore_1998} found the absorption to increase with redshift, and not with luminosity, so perhaps the absorption is a cosmological effect, not related to the jet.  One of the goals of the present work is to elucidate this ambiguity. 

\textit{XMM-Newton} with its unprecedented effective area, produced much higher signal-to-noise (S/N) X-ray spectra, and enabled the detection of even more high redshift objects, both RLQs and RQQs, that were too faint for \textit{ROSAT} \citep[e.g.,][]{Fabian_2001, Worsley_2004b, Worsley_2004a, Yuan_2005, Grupe_2006, Sambruna_2007, Saez_2011}. 
\cite{Page_2005} detected significant soft X-ray absorption towards 9 out of 16 high-$z$ RLQs using \textit{XMM-Newton} spectra. Having found no correlation between absorption and the occurrence of intervening systems along the line of sight, they concluded that the absorption was likely intrinsic, and associated with the quasar or its host galaxy. However, the 7/16 unabsorbed RLQs in the same sample pose a challenge to this interpretation.  

The puzzle of X-ray absorption of high-$z$ sources is especially intriguing given that the physics of soft X-ray absorption below 1 keV is well understood. 
Sub-keV absorption is dominated by photo-ionization of heavy elements, such as C, N, O, and Fe. 
Therefore, the total effective photo-ionization cross section 
depends on the chemical composition, and on the ionization of the absorber.
For neutral, solar composition gas, the contribution of metals dominates the cross section above 0.5~keV.
Hence, the detection of absorption at such energies provides hints to the composition and physical properties of the absorber. 
Diagnostics of absorption spectral lines could potentially help identify the mysterious nature of high-$z$ quasar absorbers.
However, contemporary X-ray grating spectrometers on board {\it XMM-Newton} and {\it Chandra}, which can resolve lines, suffer from insufficient effective area.
Thus, the time needed for extensive spectroscopic studies of a meaningful sample of high-$z$ quasars is prohibitive.
Instead, the moderate spectral resolution available with X-ray CCD detectors is used.
Indeed, the CCD spectra analyzed in the present work reveal only broad-band and not discrete features, which allow to measure the overall photo-ionization absorption and not individual lines. 

Aside from quasars, soft X-ray absorption is also detected towards gamma-ray burst (GRB) afterglows \citep[e.g.,][]{Campana_2010}. \cite{Behar_2011} suggested a diffuse intergalactic medium (IGM) dominated absorption scenario, based on measurements towards more than a hundred GRBs from the \textit{Swift} X-ray telescope, and based on the relatively constant optical depth measured for $z > 2$ GRBs. To further test this scenario, the authors analyzed 12 high S/N $z > 2$ quasar spectra from \textit{XMM-Newton}, also included in the present sample, and found that at $2 < z < 2.5$, the soft X-ray absorption of the quasars is significantly less than at $z > 2.5$.
Their sample was too small to conclusively test the IGM absorption hypothesis. Subsequent works by \citet{Campana_2012}, \citet{Watson_2013}, and \cite{Starling_2013} provide some support to the IGM conjecture, but also alternative interpretations of the X-ray absorption towards GRB afterglows.
Therefore, an additional examination of intermediate redshift ($z < 2$) quasars is required, which along with further analysis, is carried out in this work. 

Due to the limited spectral resolution, the physical state of the absorber can not be constrained even with the highest S/N X-ray CCD spectra.
In particular the redshift, the ionization (temperature), and the metallicity can only be assumed.
\cite{Behar_2011} showed that a neutral diffuse absorber with abundances of 0.2 - 0.4 solar could explain the observed absorption trend with redshift.
Recently, \cite{Starling_2013} revisited the IGM X-ray absorption hypothesis by analyzing both the \textit{Swift}/XRT GRB afterglows and a much larger sample of SDSS quasars.
They improved on the analysis of \cite{Behar_2011} by allowing ionization of the IGM absorber, which is more realistic and effectively reduces the photo-ionization cross section, thus requiring even more than the already high column density.
They find that moderately ionized IGM gas could also explain the observed absorption towards GRBs and quasars.

The present paper seeks to assemble an extensive sample of high-$z$ quasars in order to obtain a more complete picture of quasar absorption as a function of redshift and of quasar properties.
Throughout this work, we assume a Friedman cosmology with standard parameters of $H_0 = 71~\rm{km~s}^{-1} \rm{Mpc}^{-1}$, $\rm{\Omega}_{\rm{M}} = 0.27$ and $\rm{\Omega}_{\rm{\Lambda}} = 0.73$. The quasar sample is presented in section \ref{sample}, and the method of fitting and analysis is in section \ref{Method}. The results are then presented in section \ref{Results}, and discussed in section \ref{Discussion}. Finally, the main conclusions are presented in section \ref{Conclusions}.

\section{Sample}
\label{sample}

\begin{figure}			

\includegraphics[width=1.0\textwidth]{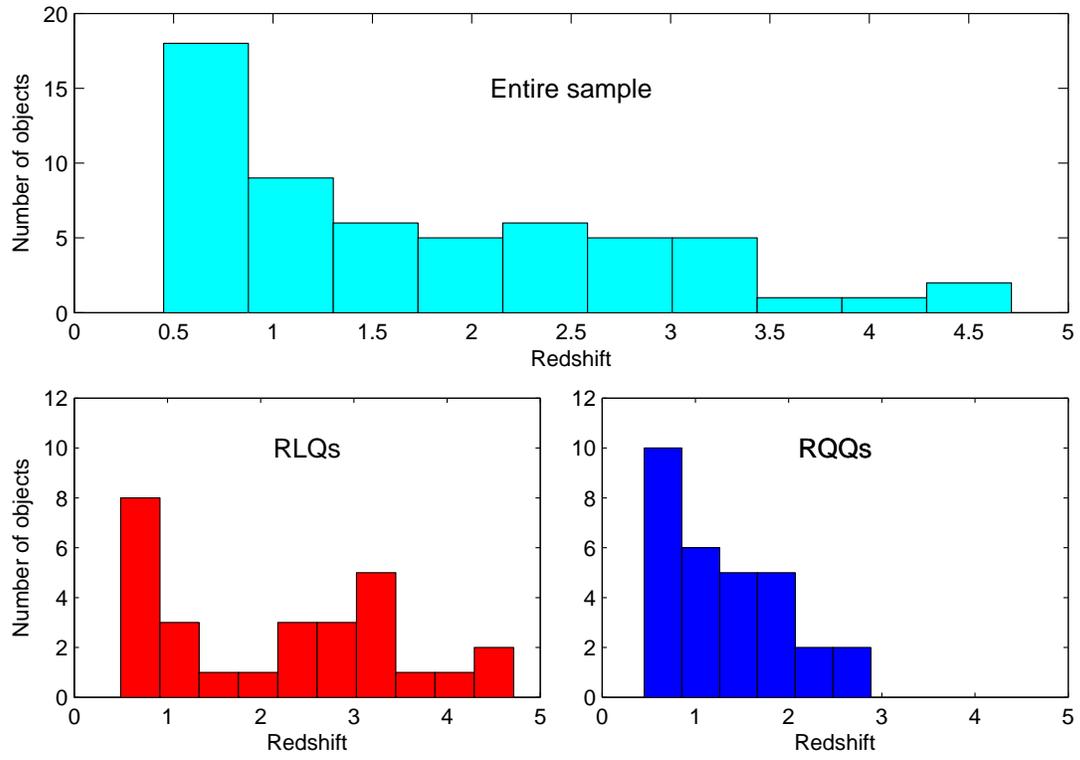}
\caption[Redshift histogram of the quasar sample]{Redshift histogram of the entire quasar sample (upper panel), radio loud quasars (lower left panel) and radio quiet quasars (lower right panel). 
}

\label{redhift_histogram_quasars_RLQ_RQQ}
\end{figure}

\begin{figure}			

\includegraphics[width=1.0\textwidth]{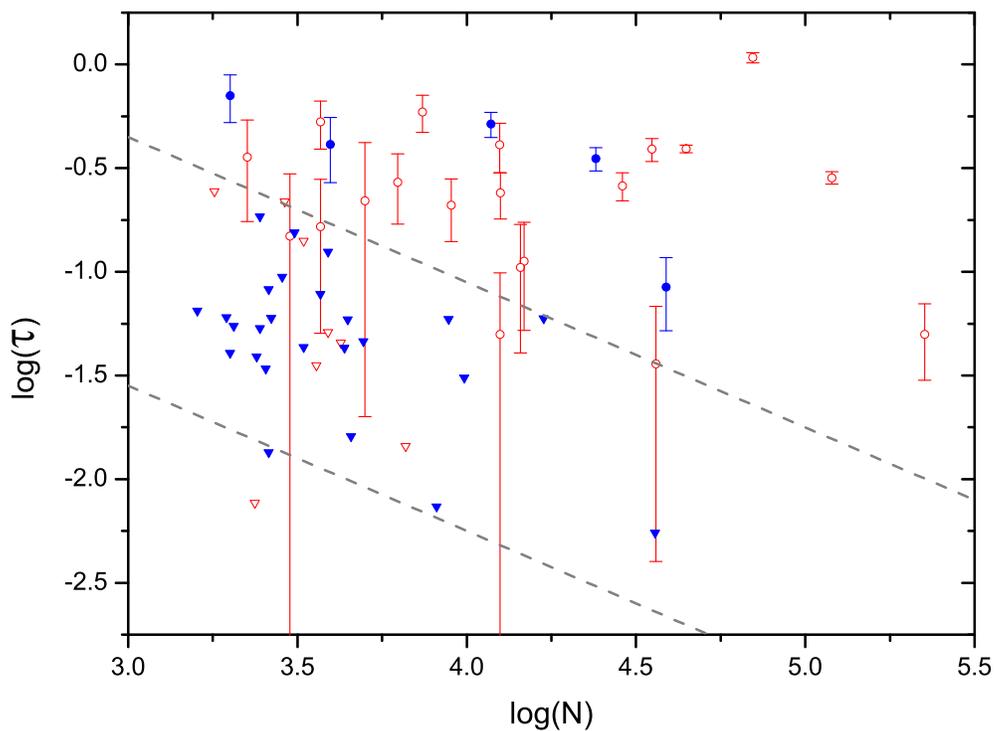}
\caption[]{Optical depth $\tau$ at 0.5~keV as a function of photon counts $N$. Circles represent quasars with statistically significant absorption, while triangles represent quasars in which only an upper limit for $\tau$ was obtained. Hollow red symbols denote RLQs, while filled blue symbols denote RQQs. The upper dashed line represents an estimated threshold 
above which absorption, if present, should be detected. The lower dashed line represents the estimated sensitivity 
down to which upper limits can be determined.
}

\label{tau_vs_photons}
\end{figure}

The present sample is X-ray selected, and consists of 58 high redshift quasars ($0.45 \leq z \leq 5$) from the \textit{XMM-Newton} archive as of 7/2012. In the analysis, we include all quasars in the archive for which the EPIC PN CCD camera on board \textit{XMM-Newton} recorded more than 1800 source photons, which ensures high S/N spectra. We use pipeline processed data products including background that was subtracted before the spectral fitting. 
The objects are listed in Table~\ref{table_of_objects}, together with their redshift, number of X-ray photons, radio properties, and the best-fit spectral index and column density. 
One could think that an archive based sample would be strongly biased and would not represent the quasar population at large. However, 23/58 objects in the present sample are not the prime target of their \textit{XMM-Newton} observation \citep[23/46, not including the highest $z$ quasars from][]{Behar_2011}, and therefore, are randomly selected.
The sample redshift histogram is presented in Figure~\ref{redhift_histogram_quasars_RLQ_RQQ} (upper panel), and demonstrates the adequate coverage of the redshift range. 
Nevertheless, the number of observed objects decreases with redshift, and there are few objects at very high $z$ ($>3.5$). In addition, RLQs (lower left panel) are distributed over a wide redshift range, while RQQs (lower right panel) do not extend beyond $z = 3$. The sensitivity of the sample to absorption is demonstrated in Figure~\ref{tau_vs_photons}. The ability to constrain optical depth at 0.5~keV as low as 0.01 can be seen. Overall we are able to measure absorption in spectra with $N$ detected photons down to $\tau \approx 56N^{-0.7}$ and upper limits down to $\tau \approx 4N^{-0.7}$.  For a few low count spectra we analyzed also the EPIC/MOS  data, which added approximately 70\% more counts,  and the upper limits decrease according to this scaling.  No upper limit, however, turns into a detection.

\begin{figure}			

\includegraphics[width=1.0\textwidth]{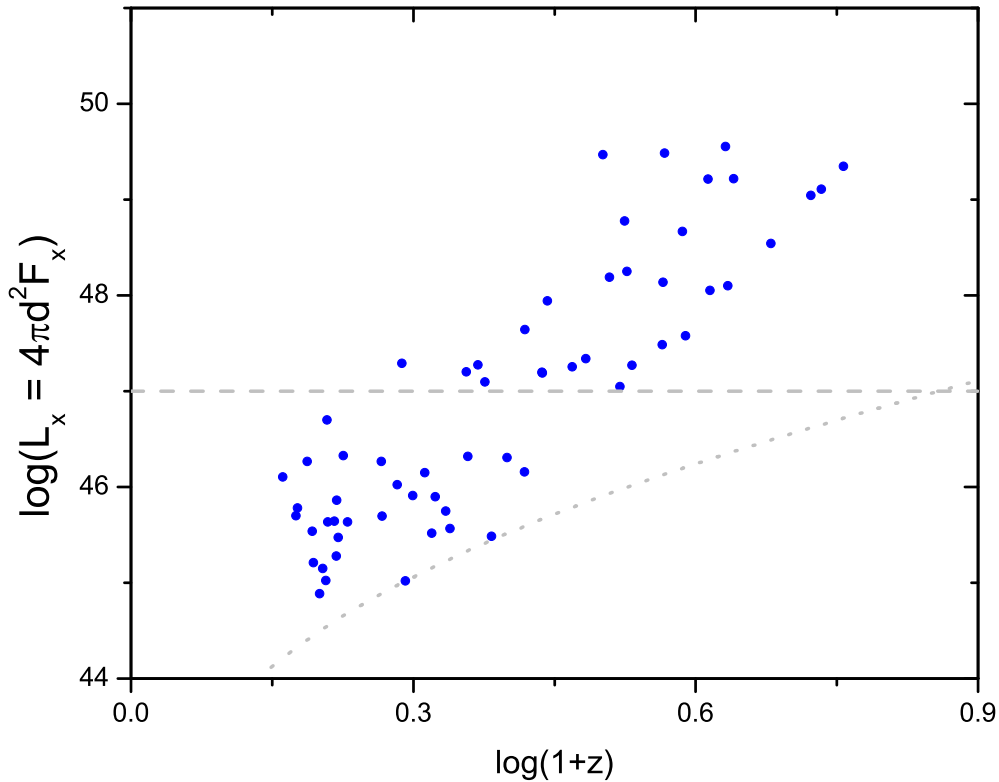}
\caption[Unabsorbed X-ray luminosity $L_x$ from 0.2 to 20~keV as a function of redshift.]{Unabsorbed X-ray luminosity $L_x$ (0.2 to 20~keV) as a function of redshift, calculated using the best fit model parameters for each source, after removing the absorption components, and assuming isotropic emission. The dotted curve represents the sample's approximate detection threshold (see text). The high luminosity values above the dashed line ($\log(L_x)=47$ ; $L_x$ in $\rm erg~s^{-1}$) are due to the isotropic emission assumption in the calculation being applied to what are likely beamed sources. The relatively tight correlation between $\log(L_x$) and $\log(1+z$) reflects the flux limited sample at high-$z$, and a limited volume effect at low-$z$. }

\label{L_x_vs_redshift}
\end{figure}

In Figure~\ref{L_x_vs_redshift} we present the unabsorbed X-ray luminosity $L_x$ between 0.2 and 20~keV
plotted against redshift. 
Luminosities are obtained from the best-fit absorbed power-law models (see section \ref{X-ray fitting} for details).
Overall, in this sample, X-ray luminosity correlates well with redshift, and a simple linear fit gives $L_x \propto (1+z)^{7.1 \pm 0.5}$.
The assumption of isotropic emission, i.e., $L_x=4\pi d^2 F_x$, where $d$ is the luminosity distance
leads in some cases, as can be seen in Figure~\ref{L_x_vs_redshift}, to high luminosity values of $L_x > 10^{48} \mathrm{erg} ~ \mathrm{s}^{-1}$, suggesting these objects are likely beamed.
The fact that these extremely bright objects are observed only at high redshift indicates that they are rare and that their beaming angle is small. 
The dashed curve represents the approximate X-ray detection threshold of the sample, which is the luminosities at which  
EPIC PN would record 2000 source photons  
within 10 ks from an unabsorbed power-law spectrum with a photon index of $\Gamma=2$. 
The sample reaches the detection threshold at low $L_x$, but at high $L_x$, due to the bimodality of the quasar population, the observed luminosity values are 1-2 orders of magnitude above the threshold. 

\begin{figure}			

\includegraphics[width=1.0\textwidth]{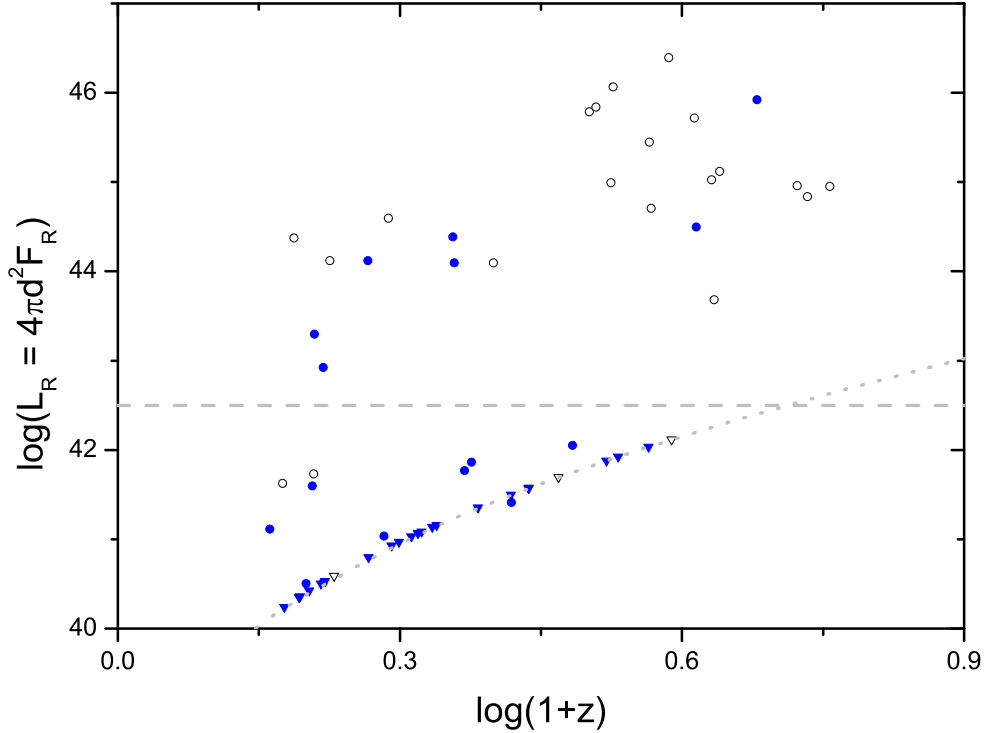}
\caption[Radio luminosity $L_R$ at 5~GHz as a function of redshift.]{Radio luminosity $L_R$ at 5~GHz as a function of redshift. Circles represent radio flux detections, while triangles represent radio flux upper limits. When radio flux detection was not available, the FIRST Survey threshold of 1~mJy (dotted curve) was taken as an upper limit. Hollow black symbols denote quasars that show significant X-ray absorption at 0.5~keV, while filled blue symbols denote quasars for which only upper limit for their optical depth was obtained. The high luminosity values above the dashed line for $\log(L_R)=42.5$ ($L_R$ in $\rm erg~s^{-1}$) are due to the isotropic emission assumption in the calculation being applied to what are likely beamed sources. Evidently, most unabsorbed objects (upper limits in $\tau$) are also fainter in the radio band (see also Figure~\ref{tau_vs_redshift_RLQ_RQQ}). 
}

\label{L_R_vs_redshift}
\end{figure}

\begin{figure}			

\includegraphics[width=1.0\textwidth]{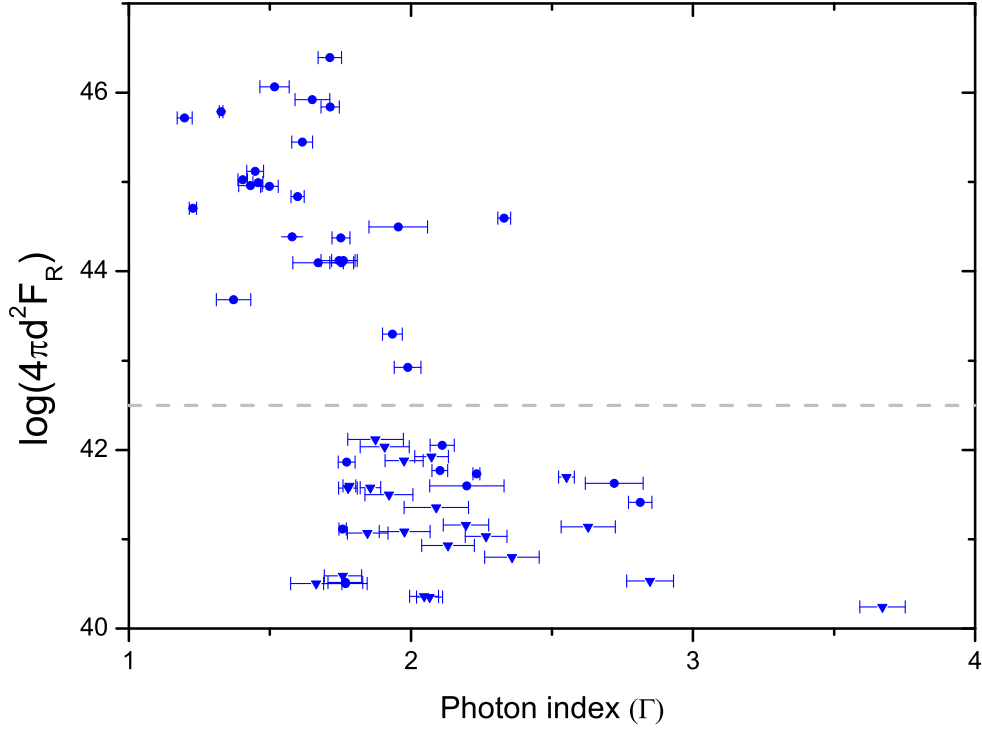}
\caption[Radio luminosity $L_R$ at 5~GHz as a function of the best fit X-ray photon index ($\Gamma$).]{Radio luminosity $L_R$ at 5~GHz as a function of the best fit X-ray photon index ($\Gamma$). Circles (triangles) denote radio detections (upper limits). Likely beamed objects (above the dashed line; $\log(L_R)>42.5$ ; $L_R$ in $\rm erg~s^{-1}$), have distinctively lower $\Gamma$ values than less luminous unbeamed objects (below the dashed line). The mean photon index value in the radio loud region is $1.64 \pm 0.26$, while in the radio quiet region it is $2.14 \pm 0.42$. 
}

\label{L_R_vs_gamma}
\end{figure}

We searched the literature for radio flux and radio loudness ($R \equiv \nu L_{\nu} (5 \mathrm{GHz}) / \nu L_{\nu} (4400 \mathrm{\AA})$) of the sample quasars. 
In Figure~\ref{L_R_vs_redshift}, we plot the radio luminosity $L_R=\nu L_\nu$ at 5~GHz, against redshift. 
For objects with no radio data in the literature, we used the FIRST Survey \cite[see][]{Becker_1995} threshold of 1~mJy as the upper limit (dotted curve). 
It turns out that 48\% of the objects in the present sample are radio loud, while generally their fraction of all quasars is much lower \cite[5-10\%, e.g.,][]{Jiang_2007}. This is a result of the sample being X-ray selected, thus, biasing it towards very high X-ray and radio luminosity. 
Objects above $\log (L_R)=42.5$ ($\rm erg~s^{-1}$ ; dashed line) have exceedingly high luminosity values, and are likely beamed. All of these are RLQs ($\log R > 1$). 
RQQs at very high redshift are not observed, since their radio luminosity is below the detection threshold.
The dichotomy of RQQs and RLQs is also manifested in the spectral slope index $\Gamma$. A plot of $L_R$ vs. $\Gamma$ is presented in Figure~\ref{L_R_vs_gamma}. The RLQs can be seen to have flatter spectra ($\Gamma_{\mathrm{median}} \approx 1.6$) compared to the RQQs ($\Gamma_{\mathrm{median}} \approx 2.0$). This result was already found by \cite{Wilkes_1987} in \textit{Einstein} IPC spectra.

\section{Method}
\label{Method}

\subsection{X-ray fitting}
\label{X-ray fitting}

As mentioned in the Introduction, as long as no absorption lines are detected, even the highest S/N CCD spectra can not distinguish between local and redshifted absorption, or indicate the chemical composition of the absorber(s). 
Since these properties, as well as the ionization, directly affect the deduced hydrogen column density $N\mathrm{_H}$, and since we have no handle on their actual values, it is more meaningful to discuss the optical depth at a given energy and not $N\mathrm{_H}$.
As long as the quality of the fit is good, and in the present work, owing to the high S/N, all spectral fits are statistically very good, the measured optical depth is independent of the model used to obtain it.
The present method can be described as follows:
The spectrum of each quasar is fitted with the \textit{XSPEC} software package\footnote{http://heasarc.gsfc.nasa.gov/docs/xanadu/xspec/}. 
The \textit{XSPEC} model used is an absorbed power-law, where the absorption is separated into its galactic and extragalactic components. 
In the fit, both are assumed to be neutral and to have solar abundances \citep{Asplund_2009}.
The galactic column density ($N\mathrm{_H^{Gal}}$) values are taken from HI
21~cm measurements \citep{Kalberla_2005}, and fixed in the fit. We do not include an $\mathrm{H}_2$ contribution, whose role was just pointed out by \citet{Willingale_2013}.
For the extragalactic component,  the absorber is fixed at the host redshift, an arbitrary choice that affects the deduced absorbing column $N\mathrm{_{H}}(z)$.
Therefore, we convert $N\mathrm{_{H}}(z)$ into optical depth at the observed energy $E$ using

\begin{equation} \label{eq:tau_eq}
\tau(E)=\sigma[(1+z)E]N\mathrm{_{H}}(z) \label{eq:tau_equation}
\vspace{0.5cm}
\end{equation}

\noindent where $\sigma[(1+z)E]$ is the total photo-ionization cross section per hydrogen atom at a photon energy of $(1+z)E$, and for neutral solar-metallicity gas. Since the cross section decreases with photon energy as $\sigma \sim E^{-2.5}$, for a given value of $\tau (E)$, the column density will depend strongly on the redshift chosen for the absorber $N\mathrm{_{H}} \sim \tau (E) (1+z)^{2.5}$.

\begin{figure}[pt!]		
 \begin{center}
 \vglue0.0cm
 {\includegraphics[angle=-90,width=5.3cm]{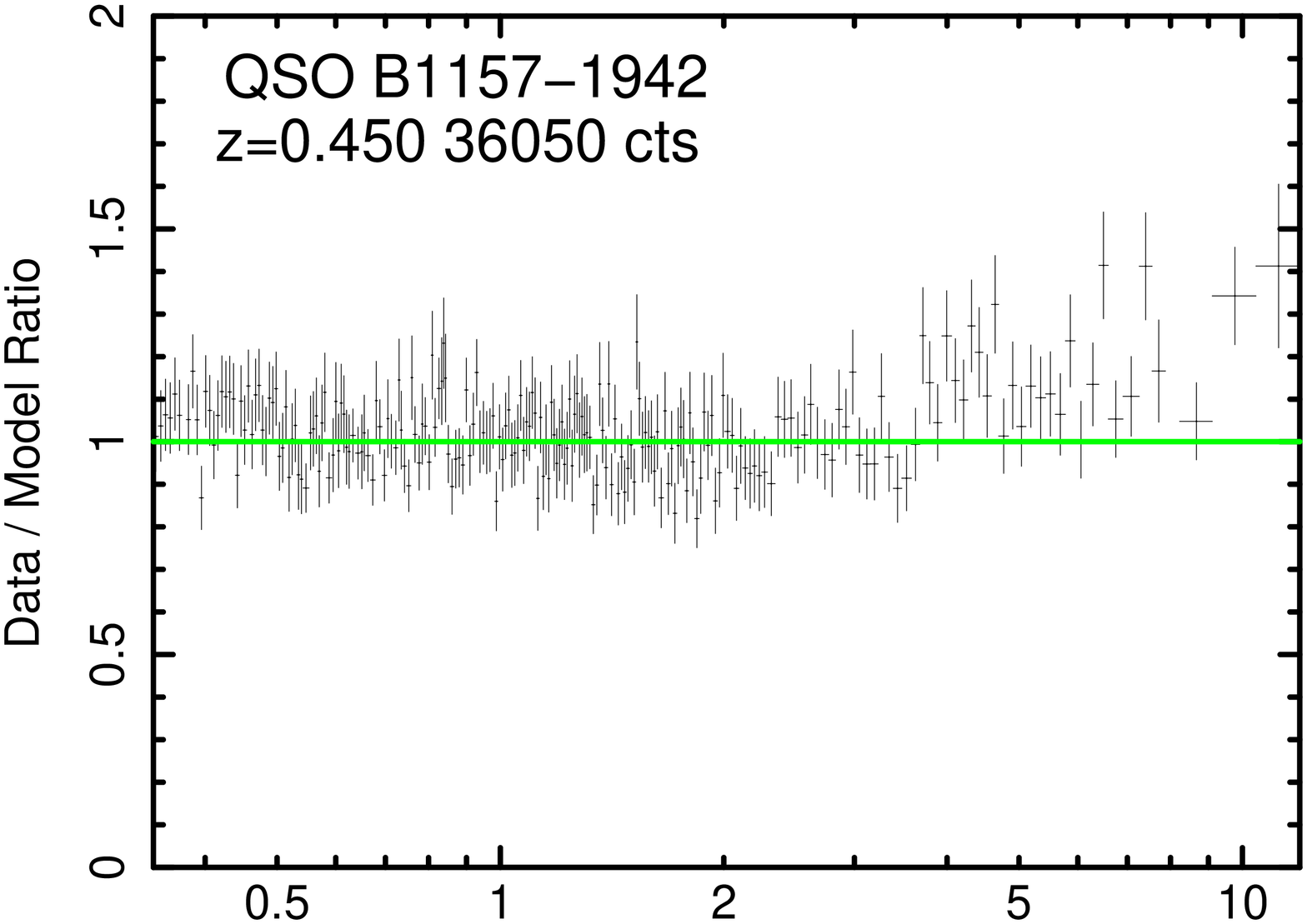}}
 {\includegraphics[angle=-90,width=5cm]{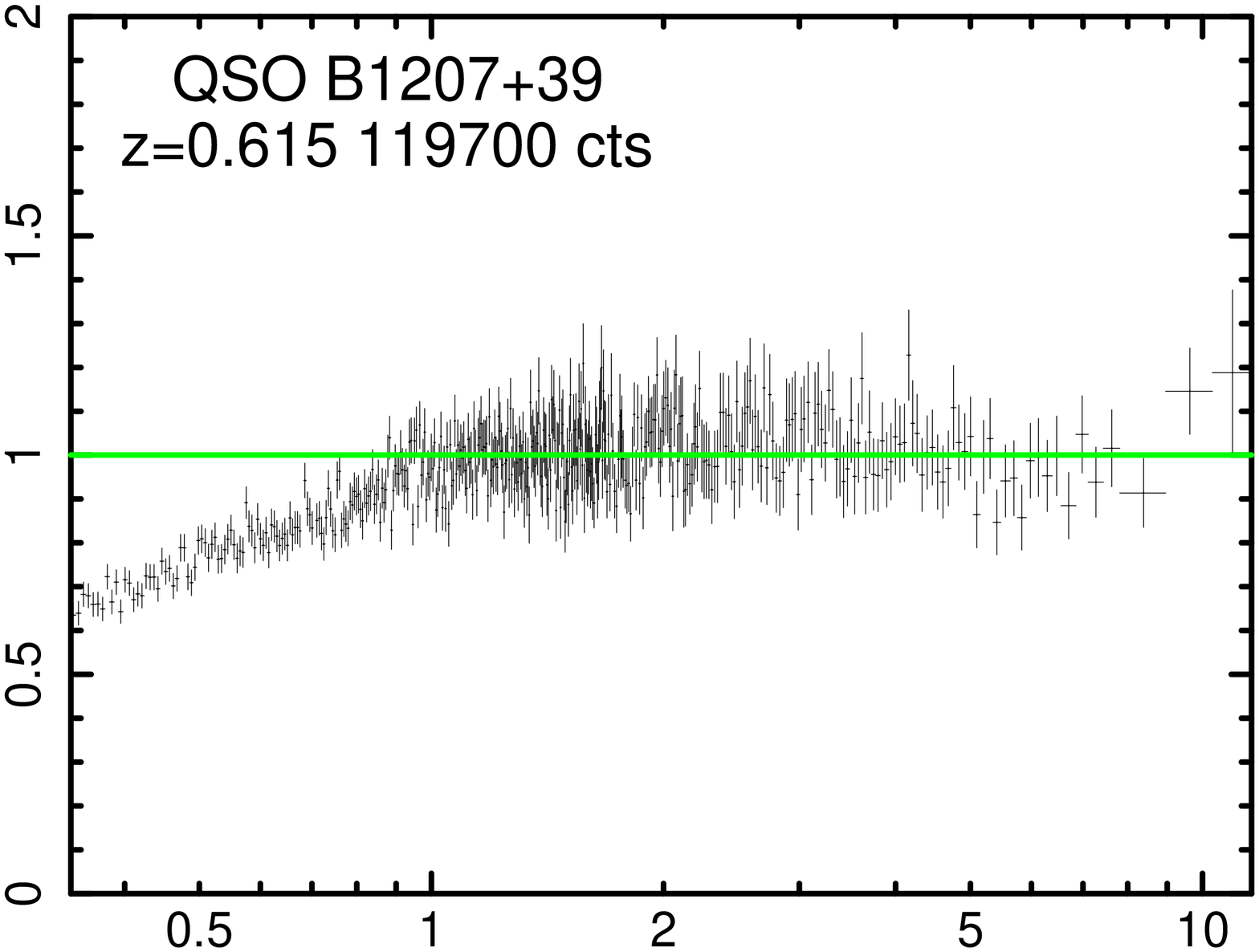}}
 {\includegraphics[angle=-90,width=5.cm]{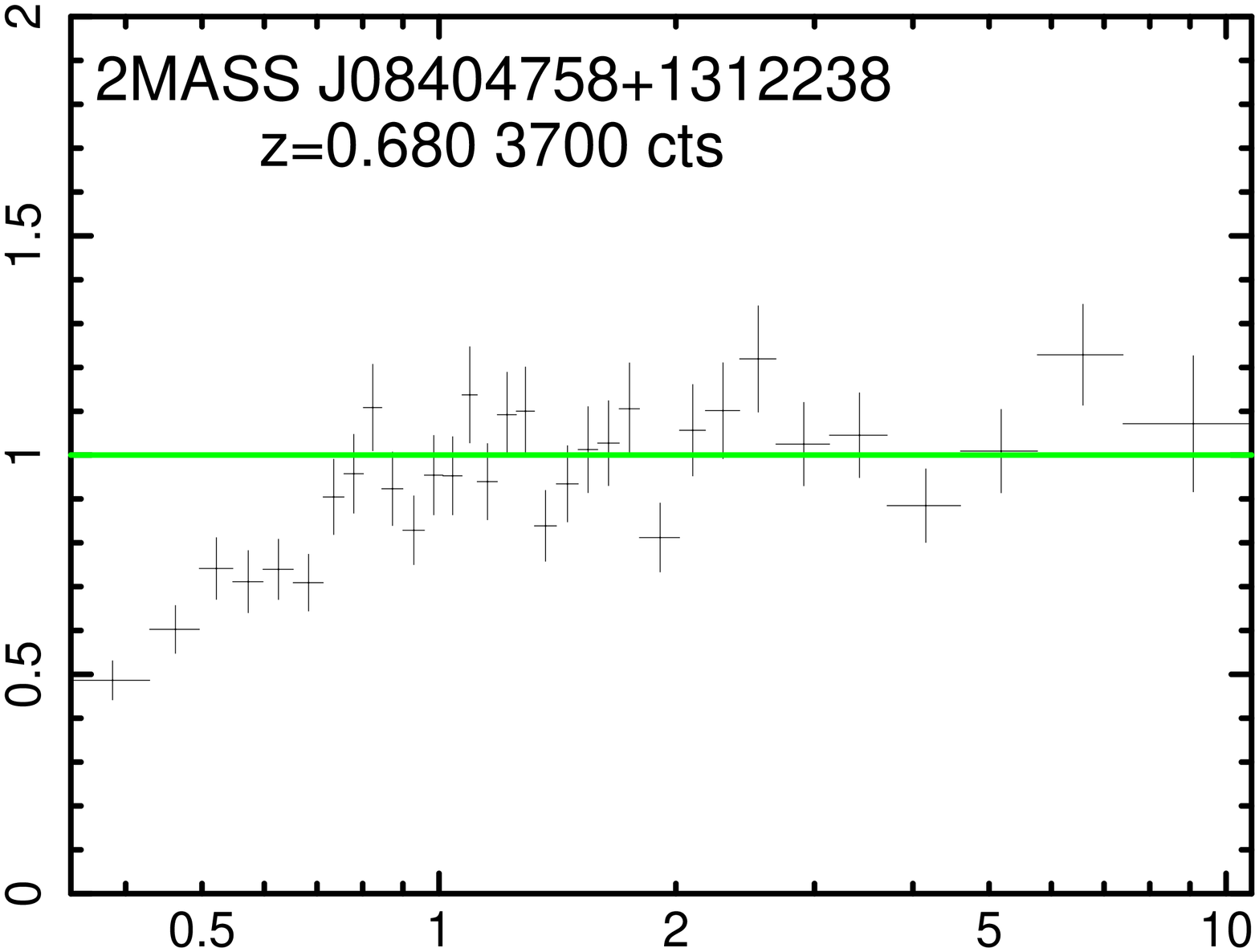}}
 {\includegraphics[angle=-90,width=5.3cm]{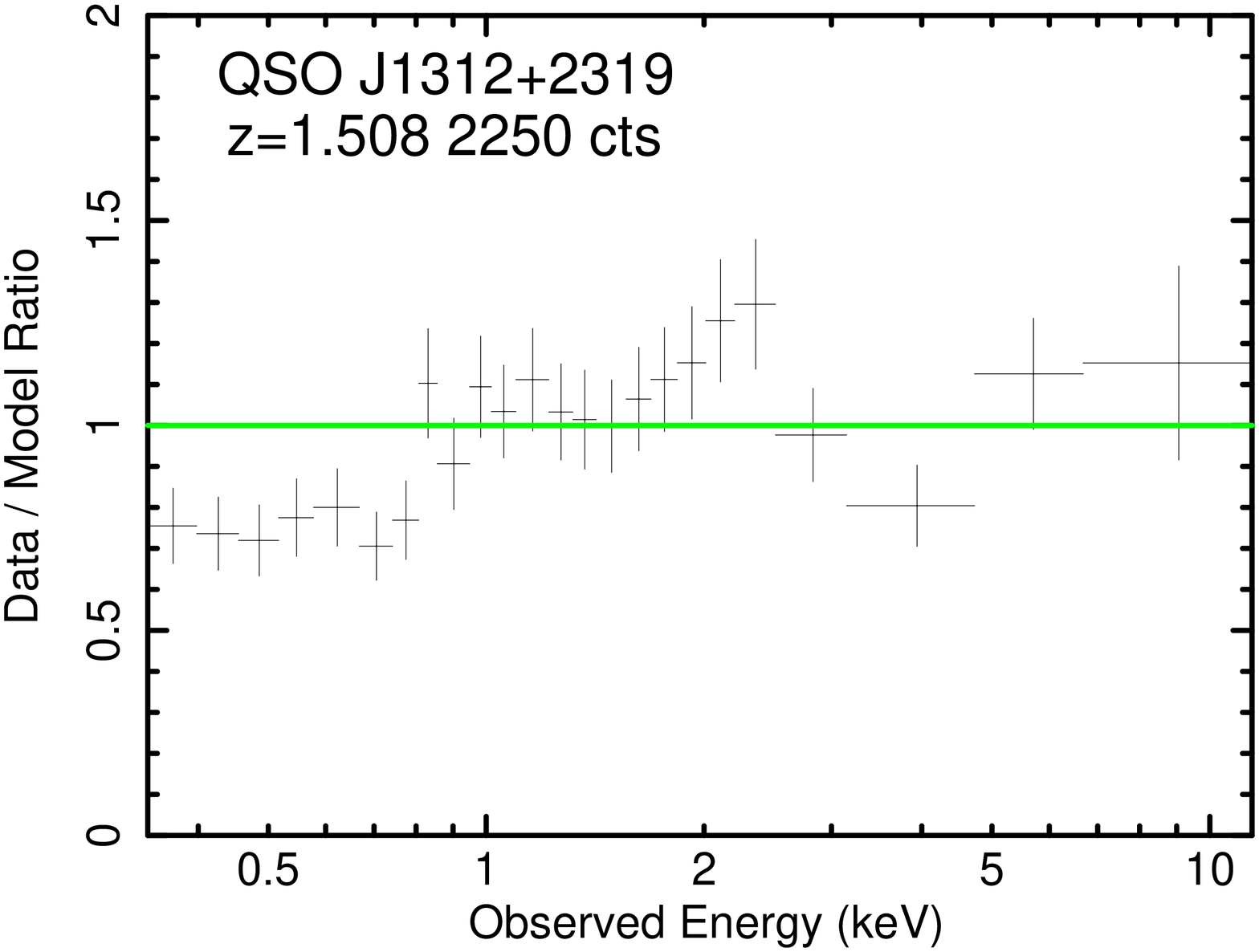}} 
 {\includegraphics[angle=-90,width=5cm]{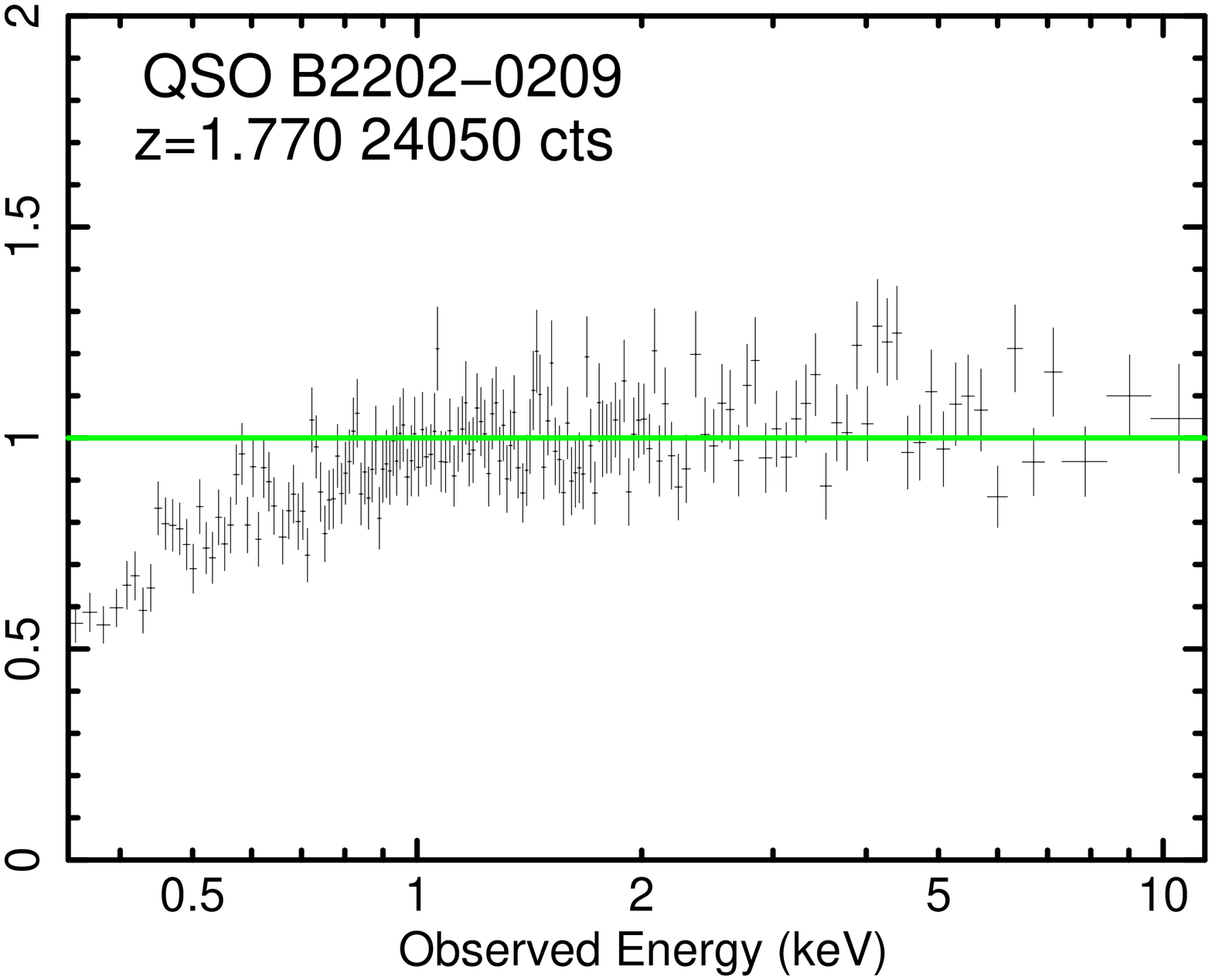}}
 {\includegraphics[angle=-90,width=5cm]{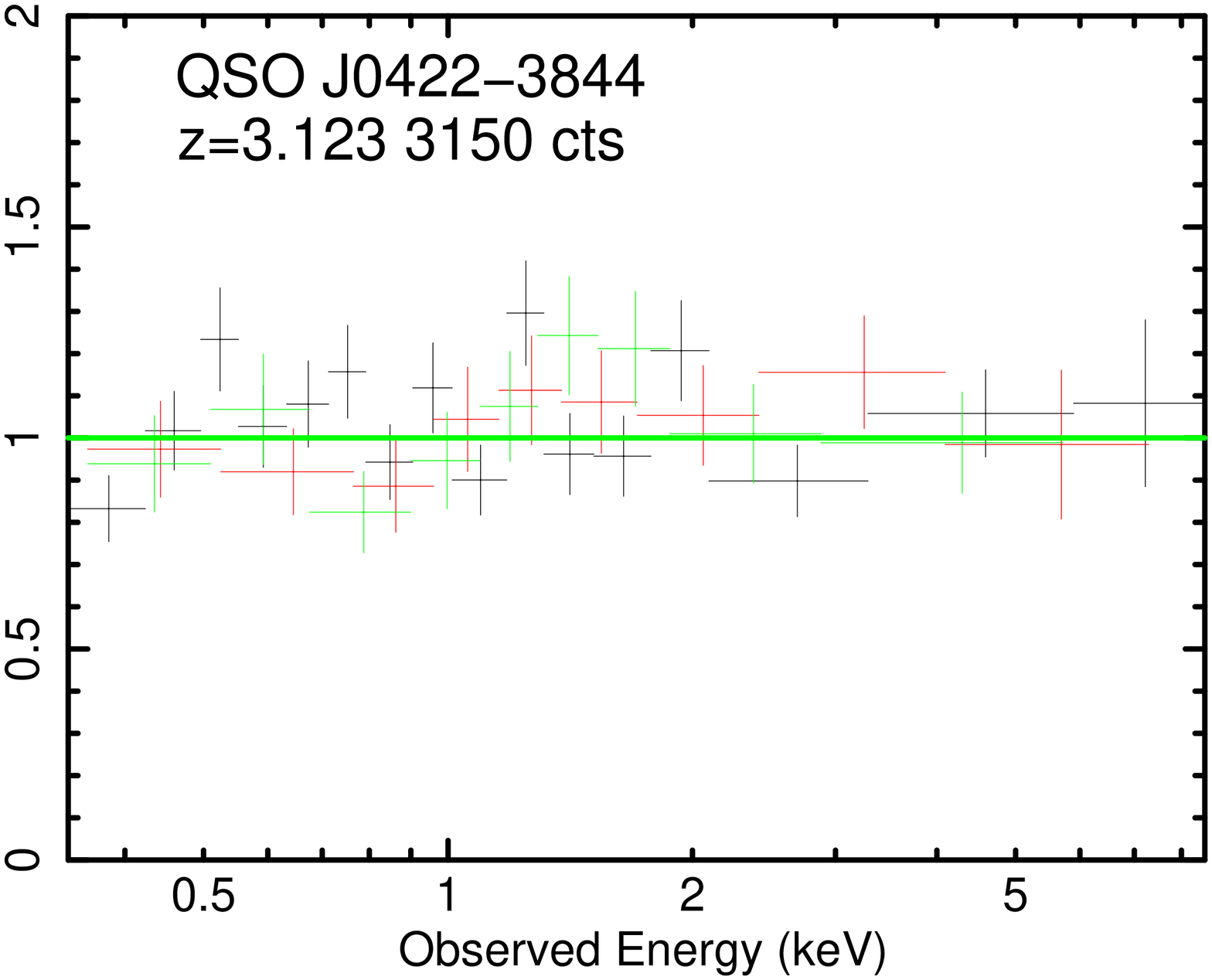}}
 \vglue0.0cm
 \caption{Data to model ratio for six selected quasars with different $\tau$/$z$/number of photons,
after the extragalactic absorption component has been removed from the model. The apparent drop of the ratio at low energies ($E < 1$~keV) reflects the additional photoelectric non-Galactic absorption toward the quasar. For QSO J0422-3844, the two EPIC/MOS spectra are also included in the fit and shown (lower right panel, in color in the electronic version).
}
 \label{data_to_model_ratio} 
 \end{center}
 \end{figure}

In Figure \ref{data_to_model_ratio} we plot the data-to-model ratio for six selected quasars, after the extragalactic absorption component has been removed from the model. Thus, the apparent turnover at low energies ($E < 1$~keV) reflects the additional photoelectric non-Galactic absorption toward the quasar. These six quasars vary in their values of $\tau$, $z$, and number of photons, and exemplify the properties of the present sample.
Other parameters obtained from the fit are the photon index ($\Gamma$), and the power-law normalization. We subsequently extrapolate the power law to estimate $L_x$ in the entire X-ray range of 0.2 - 20 keV.


\subsection{Censored statistics}
\label{censored_statistics}

Since 57\% of the objects in the present sample provide only upper limits for their optical depth, and since 40\% of the objects have no radio detection, we need to use censored statistics.
Upper limits in $\tau$ are determined by the S/N of the X-ray spectra, while upper limits in $L_R$ are taken as the detection threshold of the FIRST survey.
We employ several commonly used censored statistics methods from ASURV, the Pennsylvania State University suite of codes\footnote{http://astrostatistics.psu.edu/statcodes/asurv} \citep{Isobe_1986}, which are, basically, generalizations of regression methods and correlation tests. 

The three regression methods we use are the EM Algorithm with the Kaplan-Meier estimator, the EM Algorithm with a normal distribution, and the Schmitt regression for doubly censored data \citep{Isobe_1986}. For the Kaplan-Meier estimator function, see \cite{Feigelson_1985}. EM stands for two steps of the algorithm: expectation and maximization. First, "true" values are estimated for the censored data, according to their upper limit value. Then, using these estimated "true" values, maximum likelihood estimators of the regression coefficients are found. These two steps are then iteratively repeated, until the regression coefficients are sufficiently constrained.
These two methods treat censoring only in the dependent variable, and the difference between them is the distribution of the dependent variable values about the regression line. 
This difference, as it turned out, is not acute in our case. 
Schmitt's method, on the other hand, accounts for censoring in both dependent and independent variables, and is carried out by dividing the two-variable plane into bins. Then, a two-dimensional probability function is assigned to each bin, and by taking various moments of the function, regression coefficients can be obtained.

In addition to regression methods, we also use two correlation tests: 
Generalized Kendall's and Spearman's rank correlations, which have been modified to treat censored data
that evaluate the level of correlation between the two sets of parameters. 
Both tests account for censoring in both the dependent and independent variables, and give the significance of the correlation, as well as the null hypothesis probability. 
The results of this analysis are presented in section \ref{Results}.

\section{Results: Sample Statistics}
\label{Results}

\begin{figure}			

\includegraphics[width=1.0\textwidth]{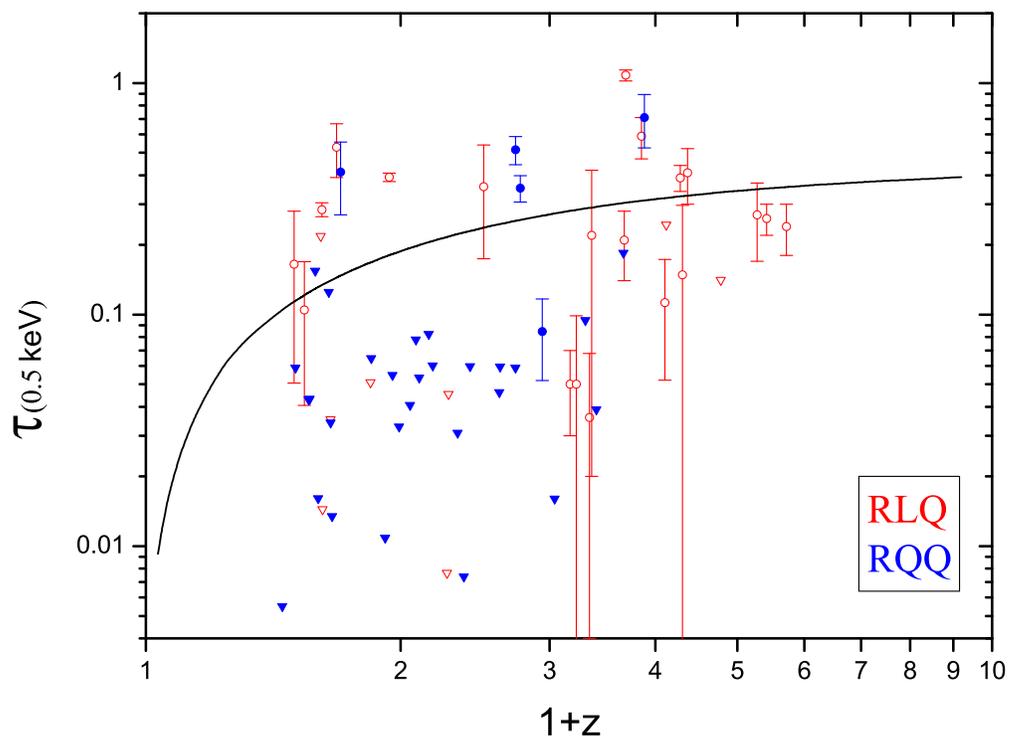}
\caption[Optical depth $\tau$ at 0.5~keV as a function of redshift.]{Optical depth $\tau$ at 0.5~keV as a function of redshift. Circles represent quasars with statistically significant absorption, while triangles represent quasars in which only an upper limit for $\tau$ was obtained. Hollow red symbols denote radio loud quasars, while filled blue symbols denote radio quiet quasars. The diffuse IGM theoretical contribution, scaled to approach $\tau=0.4$ at high $z$, is represented by the black curve \citep{Behar_2011}. While high-$z$ quasars ($z > 2$) are consistent with this curve, absorption at lower redshifts is too low. } 

\label{tau_vs_redshift_RLQ_RQQ}
\end{figure}

As described above, the model-independent absorption parameter to be measured is the optical depth of quasar spectra at 0.5~keV. The best-fit $\tau$(0.5~keV) values are listed in Table~\ref{table_of_objects}, and plotted against redshift in Figure~\ref{tau_vs_redshift_RLQ_RQQ}. In the figure, RLQs (RQQs) are denoted by hollow red (filled blue) symbols. 
Out of 58 objects in the sample, 25 show significant absorption, while the others give only upper limits for $\tau$(0.5~keV). 
Some trends can be readily seen in Figure~\ref{tau_vs_redshift_RLQ_RQQ}, but the large fraction of upper limits (57\% in $\tau$, 40\% in $L_R$) deems the analysis to remain somewhat ambiguous. This is a manifestation of the true low absorption in some of these sources, and not a result of the quality of individual spectra.
In this section, we present the soft X-ray absorption in the sample, and seek trends and correlations between the different parameters. 

\begin{figure}			

\includegraphics[width=1.0\textwidth]{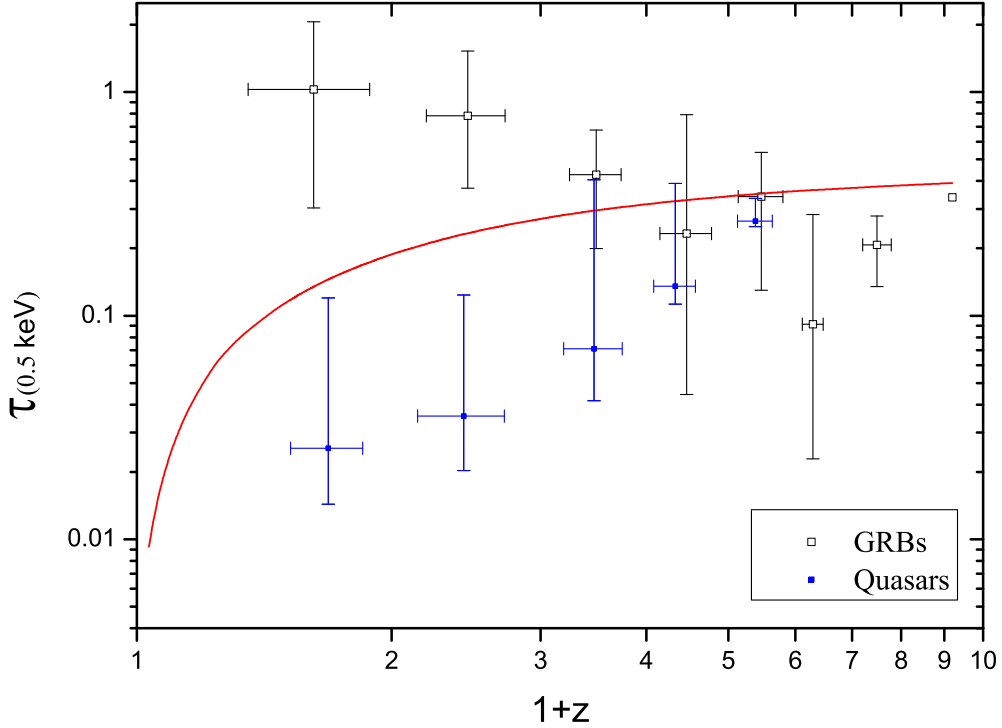}
\caption[Median optical depth ($\tau$) values at 0.5~keV over $\Delta z=1$ bins as a function of redshift (for quasars and GRBs).]{Optical depth $\tau$ at 0.5~keV as a function of redshift for the quasar (filled blue squares) and GRB (hollow black squares) samples. Data points represent median $\tau$ values calculated over $\Delta z=1$ bins. The optical depth values taken for the unabsorbed objects are half the value of their upper limit. $\tau$ error bars represent the 25th and 75th percentiles, while $1+z$ error bars represent the standard deviation in each redhift bin. The red curve represents the diffuse IGM theoretical contribution, scaled to approach $\tau=0.4$ at high $z$. Clearly, GRBs are more absorbed than quasars at lower redshifts ($z\lesssim 2$), but the two populations tend to converge at $z\gtrsim 2$. This could be due to a host contribution to the absorption in GRBs, which is not present in quasars.
}

\label{tau_median_vs_redshift_mean_quasar_GRB}
\end{figure}

Figure~\ref{tau_vs_redshift_RLQ_RQQ} features a difference between the quasars at $z > 2$ and at $z < 2$.
At $z > 2$ most objects are significantly absorbed, while at $z < 2$ most of them have only upper limits, with a few exceptions.
Although there is large scatter, there is an increase of the mean $\tau$ with $z$. Most conspicuous is the lack of unabsorbed quasars at $z > 2$.
To better demonstrate this trend, we divide the sample into redshift bins of $\Delta z=1$, and calculate the median $\tau$ in each bin. 
For the upper limits, in the lack of any better information,
we take half of the measured upper limit. 
The results are plotted in Figure~\ref{tau_median_vs_redshift_mean_quasar_GRB}, and confirm the overall optical depth increase with redshift. The Figure also compares this trend with that of GRBs, which is discussed below in section \ref{GRB sample}.

\begin{figure}			

\includegraphics[width=1.0\textwidth]{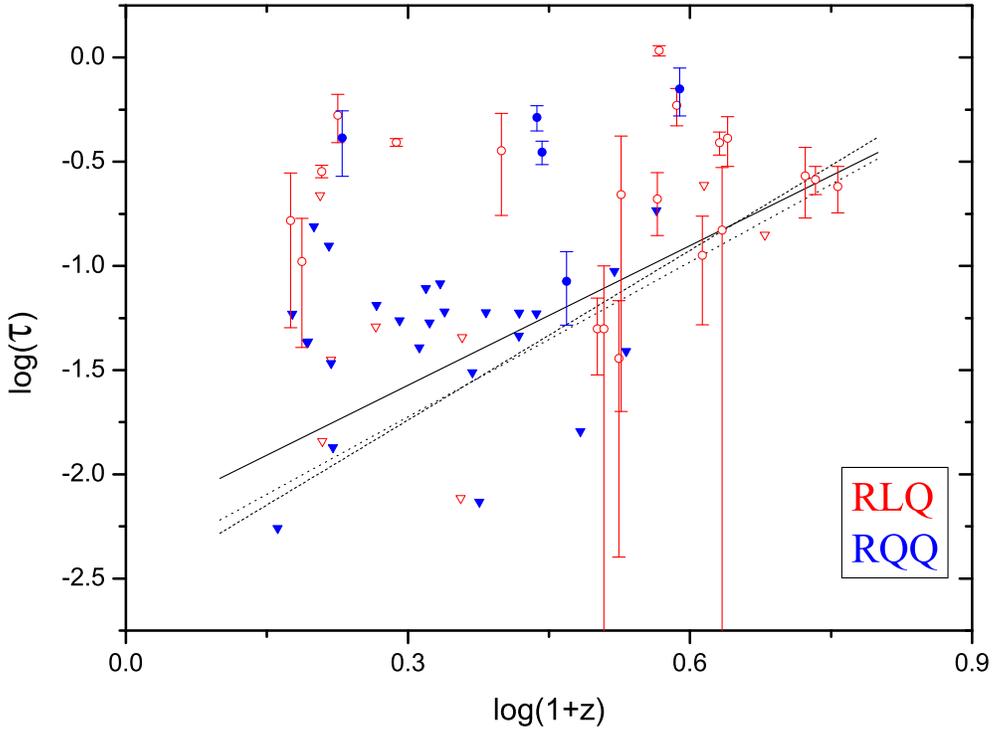}
\caption[Linear regressions of log($\tau$) vs. log($1+z$).]{Linear regression of log($\tau$) vs. log($1+z$) using three different methods implemented in the ASURV code \citep{Isobe_1986}. Solid line: the EM Algorithm with the Kaplan-Meier estimator. Dotted line: the EM Algorithm with a normal distribution. Dashed line: Schmitt's regression for doubly censored data. Circles represent absorption detections, while triangles represent $\tau$ upper limits. Radio loud (quiet) objects are in red (blue). The EM Algorithm with the Kaplan-Meier estimator (solid line) is the least restrictive method, and gives $\tau\propto (1+z)^{2.2\pm0.6}$, with a generalized standard deviation of $\sigma = 0.56$ in $\log(\tau)$.  
}

\label{tau_vs_redshift_linear_correlations}
\end{figure}

We employ censored statistics methods (section \ref{censored_statistics}) to obtain a linear regression of $\log(\tau)$ against $\log(1+z)$. 
Figure~\ref{tau_vs_redshift_linear_correlations} shows the results for three different methods \citep{Isobe_1986}, 
which give roughly the same result.
The EM algorithm with the Kaplan-Meier estimator (solid line) is the least restrictive method, and gives $\tau\propto (1+z)^{2.2\pm0.6}$ with a generalized standard deviation on $\log(\tau)$ of $\sigma = 0.56$. 
Although this result confirms a steep optical depth increase with redshift, the 57\% upper limits in the data make the astrophysical implication of the slope questionable.
In particular, there is no detection of $\log (\tau) < -1.5$ around the low-$z$ region of the regression in Figure \ref{tau_vs_redshift_linear_correlations}.
Under the assumption of a solar composition, neutral absorber, at the quasar redshift, formally, the column density scales with $z$ as $N\mathrm{_{H}} \propto (1+z)^{4.7\pm0.6}$. Given the tight correlation between $L_x$ and $z$, $L_x \propto (1+z)^{7.1\pm0.5}$ (Figure~\ref{L_x_vs_redshift}), this implies an increase of column density with luminosity of $N\mathrm{_{H}} \propto L_x^{0.7\pm0.1}$. 

\begin{figure}			

\includegraphics[width=1.0\textwidth]{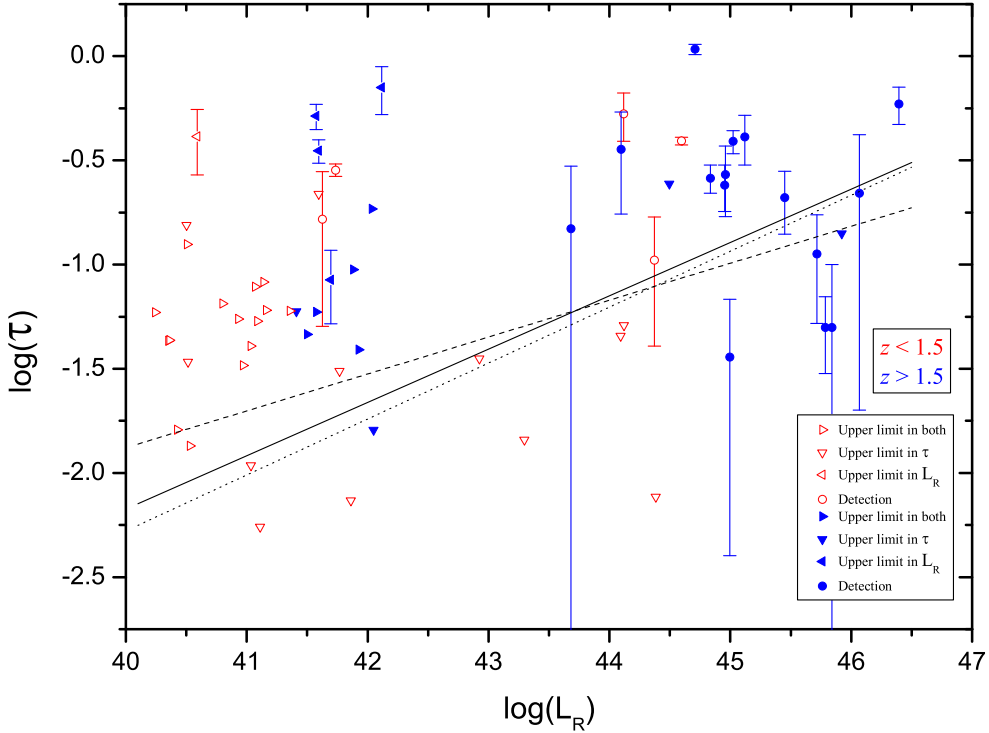}
\caption[Linear regressions of log($\tau$) vs. log($L_R$) at 5~GHz.]{Linear regression of log($\tau$) vs. log($L_R$) at 5~GHz using three different methods, implemented in the ASURV code \citep{Isobe_1986}. Solid line: Schmitt's regression for doubly censored data. Dashed line: the EM Algorithm with the Kaplan-Meier estimator. Dotted line: the EM Algorithm with a normal distribution. Detections in both variables are denoted by circles, down pointing triangles represent upper limits in $\tau$, left pointing triangles represent upper limits in $L_R$ and left-down pointing triangles represent upper limits in both variables. The $z < 1.5$ objects are in red, while the $z > 1.5$ objects are in blue. Schmitt's method (Solid line) gives the shallowest slope, because unlike the other two methods, it accounts for the censoring in both variables.  
}

\label{tau_vs_L_R_linear_correlations}
\end{figure}

Given the difference in absorption patterns detected for RLQs and RQQs \citep[e.g.,][]{Fiore_1998},
we examine the X-ray absorption dependence on radio properties in the present sample.
In Figure~\ref{tau_vs_L_R_linear_correlations}, we plot $\tau$ at 0.5~keV against radio luminosity $L_R$ at 5~GHz as well as the linear regression results. 
The bimodal distribution, i.e, the separation between RLQs and RQQs is manifested in the luminosity as $\log(L_R) \gtrsim 42.5$ for RLQs and $\log(L_R) \lesssim 42.5$ for RQQs ($L_R$ in $\rm erg~s^{-1}$).
RLQs are mostly absorbed, while RQQs give mostly upper limits.
To complicate things, RQQs in the present sample tend to be at lower redshift, while RLQs tend to be at higher redshift, so it is difficult to distinguish the redshift effect from the radio dependence. 
The Schmitt regression method for doubly censored data (Solid line) gives a shallow slope of $\tau \propto L_R^{0.2}$. Due to the heavy censoring in both parameters, however, the ASURV code cannot constrain the error on the slope, which questions its physical viability.
As in Figure~\ref{tau_vs_redshift_linear_correlations}, there is no detection of $\log (\tau) < -1.5$ around the low-$L_R$ region of the regression.
For comparison, we also plot  in Figure~\ref{tau_vs_L_R_linear_correlations} the EM Algorithm with the Kaplan-Meier estimator (dashed line), and the EM Algorithm with a normal distribution (dotted line), which we used in the previous section.
The two give roughly a similar, yet steeper than above, dependence of   $\tau$ on $L_R$, as they do not account for the censoring of $L_R$,
and thus overestimate $\tau$ in the $\log(L_R)<42.5$ region. 
Overall, all three methods suggest only a mild, statistically insignificant, optical depth increase with radio luminosity.

We also examined the dependence of $\tau$ on the radio loudness $R$. 
The results are very similar to those obtained for the radio luminosity, indicating that for the present purpose, $L_R$ is an adequate proxy for radio loudness. 
Moreover, the regression methods give roughly the same slopes as for $L_R$ (Figure~\ref{tau_vs_L_R_linear_correlations}), namely $\tau \propto R^{0.2}$ according to Schmitt's method.
Nevertheless, considering the many upper limits in optical depth and in radio luminosity, it would be advisable to repeat this analysis in the future, with deeper radio surveys.


In addition to linear regression, we also performed two correlation tests (see section \ref{censored_statistics}) between the optical depth and the parameters $z$, $R$ and $L_R$. The results of these tests are listed in Table~\ref{table_of_correlation_tests}. The correlations are unremarkable, with chance probabilities between 0.1\% and 1\%. Evidently, $\tau$ correlates best with $L_R$ according to Spearman's test, and best with $z$ according to Kendall's test, but the differences between the numbers are insignificant. This is likely a result of the biases in our sample and the correlation between redshift and luminosity, mostly $L_x$, but also $L_R$.


Given the low resolution of the spectra, it is possible that soft X-ray extragalactic absorption is being confused with an intrinsically curved spectrum that has absolutely nothing to do with photoelectric absorption.
A broken power law is expected from beamed sources \citep[e.g.,][]{Ghisellini_1998}, as a result of relativistic-particle cooling.
This possibility has been explored, e.g., by \citet{Tavecchio_2007} for RBS\,315 with inconclusive results.
%
Unfortunately, CCD spectra alone, even those with the best S/N are incapable of distinguishing between genuine absorption and a broken power-law.
Nonetheless, we fitted a broken power-law model to all of the objects in the sample that show significant extragalactic absorption. 
In the fit, the Galactic column density is fixed.
The results are listed in Table~\ref{table_of_break_energy}. 
Indeed, the reduced $\chi^2$ values are close to 1.0 and very similar to those obtained for the absorbed power-law model, as can be seen in the two last columns of Table~\ref{table_of_break_energy}.

\begin{figure}			

\includegraphics[width=1.0\textwidth]{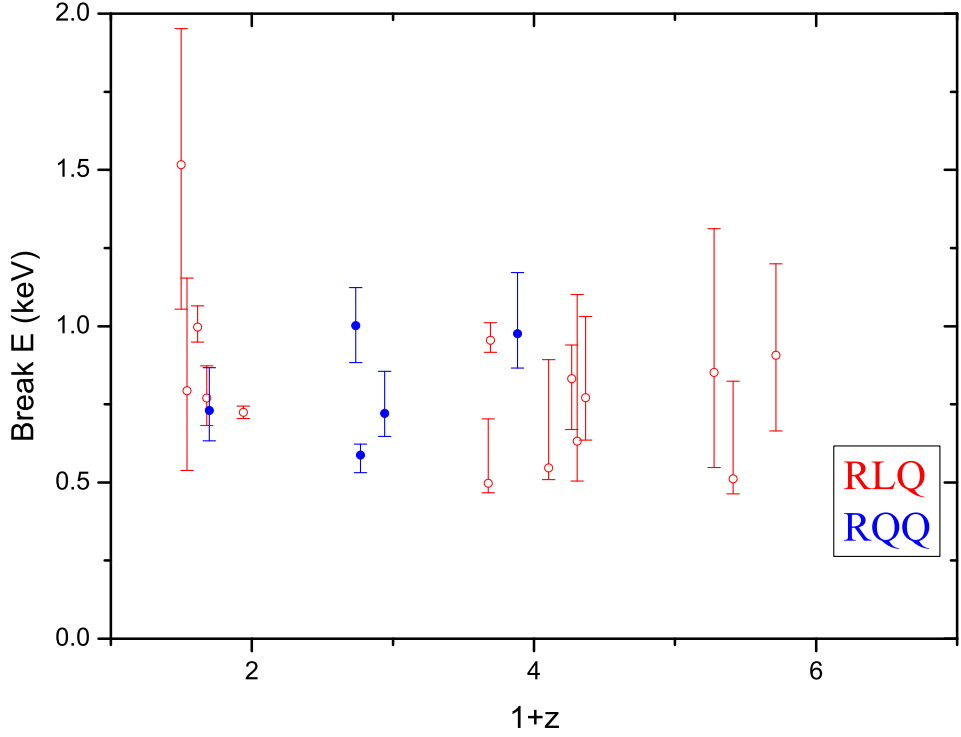}
\caption[Best fitted break energy in observed rest frame as a function of redshift.]{Best fitted break energy in observed rest frame vs. redshift. Red (blue) circles denote radio loud (quiet) quasars. Three objects with break values greater than 2.5~keV, and with barely detected absorption (break) fall outside the plot (see Tables~\ref{table_of_objects},~\ref{table_of_break_energy}). However, including these objects, a linear fit gives $E_{\rm Break}=(0.06\pm0.08)(1+z)+(0.65\pm0.19)$, which is consistent with a constant break energy, i.e, there is no dependence of the break energy on the redshift.  
}

\label{break_E_vs_redshift}
\end{figure}

Consequently, in order to obtain an idea of the viability of the two interpretations, one needs to allude to the sample statistics. 
In Figure~\ref{break_E_vs_redshift} we plot the best fitted break energy values, in the observed frame of reference, against $1+z$. 
Evidently, there is no obvious dependence.
A linear fit (including all 22 objects in Table~\ref{table_of_break_energy}) gives $E_{\rm Break} = (0.06 \pm 0.08)(1+z)+(0.65 \pm 0.19)$, which is consistent with a constant observed break energy.
In addition, we see that RQQs (filled blue circles) yield roughly the same break energies as RLQs (hollow red circles), and are also spread over a wide redshift range ($0.5 < z < 3$).
This is an interesting result that will be further discussed below in Section \ref{Discussion}. 
Note that three outliers out of 22 yield break values greater than 2.5~keV (see Table~\ref{table_of_break_energy}), which is outside the range of Figure~\ref{break_E_vs_redshift}.
We deem these targets insignificant since their absorption (or break) is barely detected.
Their $\tau$ values are $0.05 \pm 0.02$, $0.036 \pm 0.032$, and $0.22 \pm 0.21$ (Table~\ref{table_of_objects}),  
or a change in slope of $\Delta \Gamma=\Gamma_\mathrm{high}-\Gamma_\mathrm{low} = -1.17\pm0.66$, $-0.10\pm0.06$, and $-0.39\pm0.27$ respectively (Table~\ref{table_of_break_energy}).
The concave spectrum and relatively large errors likely imply these are insignificant breaks that can be dismissed.

\section{Discussion}
\label{Discussion}

\subsection{GRB sample}
\label{GRB sample}

\begin{figure}			

\includegraphics[width=1.0\textwidth]{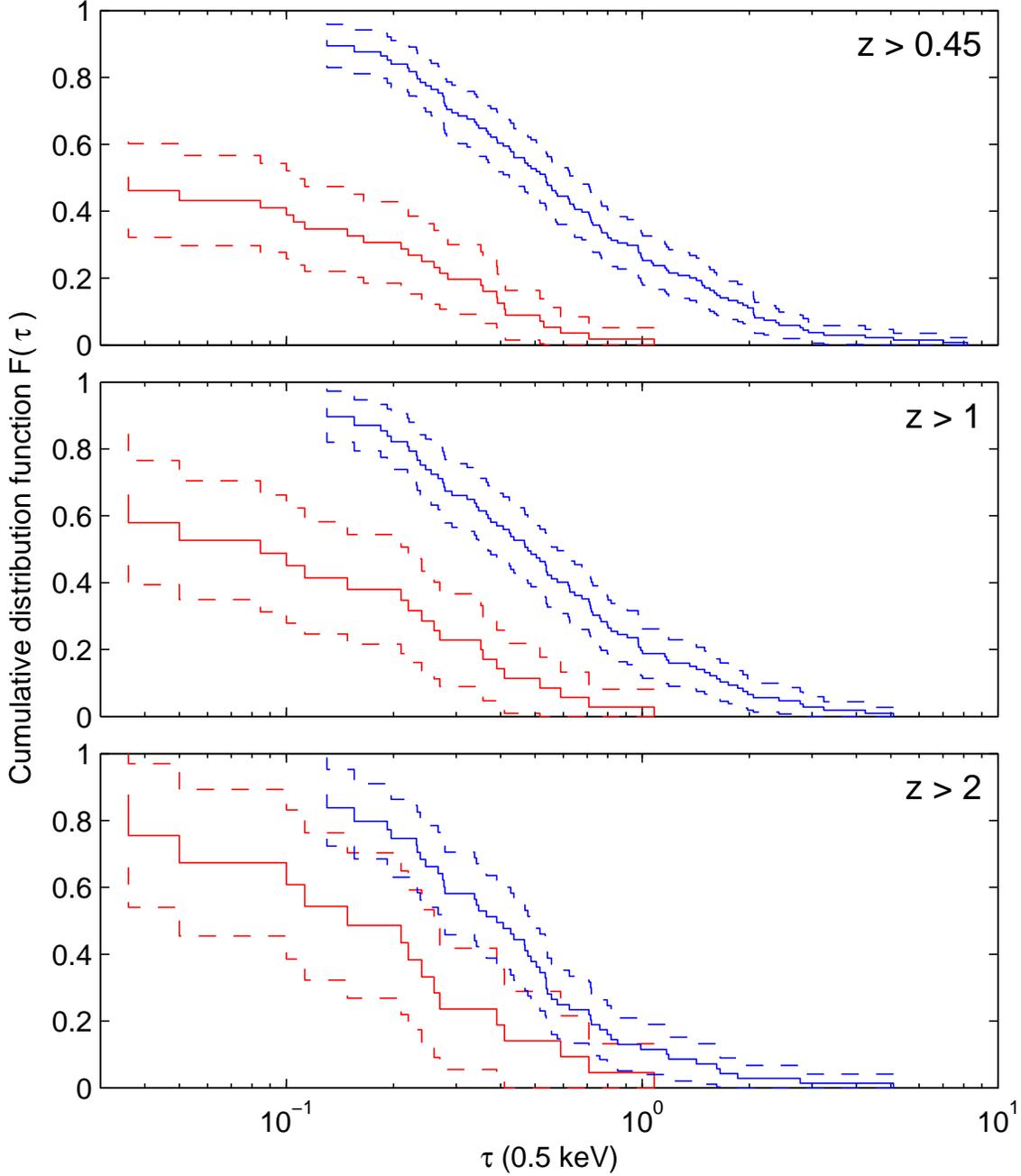}
\caption[Cumulative distribution functions $F(\tau)$ for quasars and GRBs.]{Cumulative distribution functions $F(\tau)$ for $z > 0.45$ (top), $z > 1$ (middle) and $z > 2$ (bottom) quasars and GRBs, calculated using survival analysis methods \citep{Feigelson_1985}. Red (blue) solid lines represent quasar (GRB) distribution, while dashed lines (both red and blue) represent the $\pm 1\sigma$ confidence region. The broader uncertainty for the quasar distribution reflects the many upper limits in that sample. Clearly, the two distributions come closer to each other at higher redshifts, but generally, GRBs are significantly more absorbed than quasars.
}

\label{cumulative_distribution_quasar_GRB}
\end{figure}

An interesting reference sample, to be compared with the present quasar sample, is that of gamma-ray burst (GRB) afterglows, which are another probe of soft X-ray absorption at high-$z$. 
In Figure~\ref{cumulative_distribution_quasar_GRB} we plot the cumulative distribution function $F(\tau)$ for the presently measured optical depth at 0.5~keV, compared with the distribution of the \textit{Swift} GRB sample taken from \citet{Behar_2011}.   The physical model used for the GRB spectra is the same as in this work. 
The plot is divided into three sub-samples: $z > 0.45$ (top panel), $z > 1$ (middle panel) and $z > 2$ (bottom panel). 
While the GRB absorption distribution does not change much with $z$, that of the quasars increases with $z$, and  approaches the GRB distribution only at high-$z$.
It can be seen that, generally, GRBs are significantly more absorbed than quasars, except at $z > 2$. Note that there was no restriction on the GRB photon counts, as opposed to the current quasar sample, which is limited to high photon counts.  This reinforces our conclusion that GRB afterglows at low-$z$ are more absorbed on average than the quasars. 

The same effect is shown in Figure~\ref{tau_median_vs_redshift_mean_quasar_GRB}, where median $\tau$ values are plotted in $\Delta z=1$ bins for the quasar and GRB samples.
The cosmological mean, diffuse IGM contribution (solid curve), invoked by \citet{Behar_2011} and scaled to approach $\tau=0.4$ at high $z$, is also plotted for comparison, and is further discussed below in Section \ref{Diffuse IGM absorption}. 
The same effect is seen, where the absorption pattern of quasars is different from that of GRBs, but the two populations tend to converge to $\tau \approx 0.4$ around $z  \gtrsim  2$. 

\subsection{Broken power-law}
\label{Broken power-law discussion}

As described in Section \ref{Results}, the X-ray spectra of the quasars 
are fitted equally well with an alternative model of a broken power law, with only galactic absorption. 
There are several models which predict a power-law break in the X-ray region of the spectrum of a jet \citep[see][]{Sikora_1997}. 
However, all these scenarios depend on the physical parameters of the jet, such as the bulk Lorentz factor, electron injection function and the magnetic field. 
If the break in the power-law was intrinsic to the jet (i.e., the quasar), we would expect a trend of break energy with $z$: $E_{\rm Break} \propto (1+z)^{-1}$, perhaps with some scatter due to jet parameters.
The break energy, however, in the present sample is essentially independent of $z$ (Figure~\ref{break_E_vs_redshift}).
Moreover, the RQQ spectra turn over at the same break energy as the jetted RLQs. 
RQQs are not expected to have jets, nor a spectral break. 
We take this result as strong evidence for the curvature in the spectra being due to absorption, and not being intrinsic.
On a related note, a recent work by \citet{Furniss_2013} finds that the presence of CO emitting molecular gas in local blazars can be associated with X-ray spectral curvature: another indication that the spectra are absorbed and not intrinsically curved.

\subsection{Radio dependence versus redshift dependence}
\label{Radio dependence versus redshift dependence}

As described in section \ref{Results}, we find that the optical depth at 0.5~keV increases with $z$, as well as with radio luminosity $L_R$ and radio loudness $R$. However, the linear fits (Figures~\ref{tau_vs_redshift_linear_correlations},~\ref{tau_vs_L_R_linear_correlations}), as well as the correlation tests (Table~\ref{table_of_correlation_tests}), cannot determine which parameter is more correlated with $\tau$. 
This ambiguity is due to the heavy censoring in the present sample, even if many upper limits are tightly constrained, which considerably limits the effectiveness of the statistical tools. Furthermore, it reflects the selection biases in the sample. 
Figure~\ref{L_R_vs_redshift} in section \ref{sample} shows that very luminous objects (RLQs) are mostly observed at very high redshift ($z > 2$). In addition, RQQs at the same redshift cannot be observed due to the radio bimodality and the detection limit. 
As a result, high luminosity in the sample is correlated with high redshift, and the dependency of $\tau$ on $z$ is observationally coupled to its dependency on $L_R$.
Interestingly, an analysis similar to the present one, but with an optically selected sample of quasars by \cite{Starling_2013} shows a strong increase of column density with redshift. Their results strengthen the confidence in a genuine increase of absorption with redshift.

In the local universe, the picture is quite opposite, as most ionized absorbers are found in RQQs, except for a few detections of outflows in RLQs \citep{Reeves_2009, Reeves_2010, Torresi_2010}. However, these sources are FR II lobe dominated RLQs, and their X-rays are not attributed to a jet along the line of sight. The X-rays from the extremely luminous and radio loud high-$z$ sources in the present sample are quite different in nature. 
The connection between radio loudness and X-ray absorption, if there is one, remains to be elucidated. 
  
\subsection{Strongly absorbed radio quiet quasars}
\label{The few strongly absorbed radio quiet quasars}

Only 5/30 radio quiet quasars in the present sample show significant absorption. 
The two most absorbed radio quiet objects, with $\tau > 0.5$, are QSO B1115+080A ($z = 1.736$), and QSO B1524+517 ($z = 2.883$). 
These two objects are reported to be broad absorption line (BAL) quasars \citep{Streblyanska_2010}, namely their UV/optical blueshifted absorption lines are ascribed to energetic outflows.
Such ejection of metal-rich gas could, in principle, explain the soft X-ray absorption observed in these objects. 
The weakly absorbed RQQ, SDSS J110449.13+381811.6 ($z = 1.942$) is 
not a BAL quasar according to the common criterion of at least 10\% attenuation of the UV continuum by the absorption lines \citep{Gibson_2009}. 
Its optical depth  at 0.5~keV of $\tau < 0.1$ may indicate it hosts a weaker outflow.
Conversely, for the two remaining RQQs, RX J111750.5+075712 ($z = 0.698$) and QSO B2202-0209 ($z = 1.770$), that are rather highly absorbed with $\tau \approx 0.4$, we could not find in the literature any report of optical/UV absorption features.

Only a small fraction \citep[$\sim$10\%, but see][]{Trump_2006} of quasars are known as BAL quasars.  Moreover, BALs are equally likely to occur in RQQs and in RLQs, except for the most radio loud sources \citep{Becker_2001}, where we see the most absorption (Fig. 8). Therefore, BAL outflows are unlikely to be the general explanation for the absorption observed in our sample. 

In the local universe, there is an established kinematic connection between X-ray and UV outflows in Seyfert galaxies, although the X-ray absorbing gas carries by far more mass and energy. When it comes to BAL quasars, however, the connection between BAL properties and X-ray absorption is murkier.
\citet{Gallagher_2006}, who examined 35 BAL quasars using \textit{Chandra} spectra, did not find evidence for correlations between X-ray weakness, presumably absorption, and UV absorption-line properties. Although \citet{Streblyanska_2010} did not find any dependence between the neutral absorption measured in X-rays and the equivalent width of C {\scriptsize{IV}} in the UV, they did find an apparent trend of increasing ionized X-ray column with C {\scriptsize{IV}} absorption.
Considering the small number of five absorbed RQQs, of which two are confirmed BAL quasars, 
it is not possible to draw solid conclusions from the present sample regarding the correlation between X-ray and UV absorption.

\subsection{Clumpy and Ionized IGM}
\label{Diffuse IGM absorption}

\citet{Behar_2011} showed that a mean, diffuse IGM absorber could produce an increase of X-ray optical depth with $z$ that would tend to a constant $\tau \approx 0.4$ at $z > 2$ (see solid curves in Figures~\ref{tau_vs_redshift_RLQ_RQQ}, \ref{tau_median_vs_redshift_mean_quasar_GRB}). 
This behavior was invoked to explain why $\tau$ at 0.5~keV in high-$z$ GRBs approaches 0.4, and could arise if the known baryon content of the universe was uniformly distributed across the entire IGM.
The asymptotic value depends on the IGM metallicity and ionization. 
While at $z  \gtrsim  2$ the quasar and GRB absorption pattern do indeed tend to $\tau \approx 0.4$, at $z  \lesssim  2$ the observed absorption of many quasars is significantly lower than that expected from the naive diffuse, neutral, IGM scenario.
In order for the IGM to explain the observed quasar absorption pattern, it would need to be partially ionized, as also found by \cite{Starling_2013} (and as expected in the Warm Hot Intergalactic Medium). Moreover, the IGM is likely clumpy, to the point that most lines of sight ($27/37 \approx 70\%$ in the present sample) are relatively clear up to $z \approx 2$.

The picture changes for the highest-$z$ quasars and GRBs, where a dominant effect of cold IGM absorption is reasonably expected {\it a la} the Ly$\alpha$ forest, and is consistent with the present data.
In Figure~\ref{tau_vs_redshift_RLQ_RQQ}, a tentative increase in absorption can be seen around $z = 2.5$. 
Objects at higher redshift show stronger absorption than objects at lower redshift, with a few exceptions. 
This change can be associated with the re-ionization of He {\scriptsize{II}} in the IGM, which UV spectroscopic observations indicate occurred at $z = 2.7 \pm 0.2$ \citep{Shull_2010}. 
The ionization energy of He {\scriptsize{II}} is comparable to that of O {\scriptsize{III}} and C {\scriptsize{III}}, which could be responsible for part of the X-ray absorption. 
Therefore, the observed quasar absorption pattern might be a trace of this cosmological event.
However, at $z < 2$,  there are 10 absorbed objects (out of 37), which remain unexplained according to this interpretation, and again lend to the notion of a clumpy IGM or a host contribution.

\section{Conclusions}
\label{Conclusions}

We have studied the high S/N X-ray spectra of 58 high redshift quasars, 28 radio loud (RLQs) and 30 radio quiet (RQQs). 
While 43\% of the quasars show significant absorption at 0.5~keV, we were able to obtain only upper limits for the optical depth of the remaining 57\%. Overall, we have established the following:

\begin{enumerate}
\item 
Optical depth $\tau$ at 0.5~keV measured towards high redshift quasars seems to increase with $z$, roughly as $\tau \sim (1+z)^{2.2\pm0.6}$, and with $L_R$, roughly as $\tau \sim L_R^{0.2}$. However, given the many upper limits in $\tau$ and in $L_R$, the exact scaling in these relations should be treated with caution.
\item
Significantly more RLQs are absorbed than RQQs.
This confirms similar previous results dating back to \textit{ROSAT}, 
and might be due to a host (jet) contribution which is not present in RQQs, 
but would require a model for the location of the photoelectric absorbing atoms. 
\item
Although quasars and GRB afterglows show roughly the same $\tau$(0.5~keV) at $z > 2.5$, at lower $z$ GRBs are significantly more absorbed.
The emerging picture is that an ionized and clumpy IGM can contribute to the X-ray absorption along some lines of sight up to $z \approx 2.5$, while less ionized IGM dominates the X-ray absorption to most lines of sight at higher redshifts.
\item
The uniform break energy, independent of $z$, strongly suggests that the curvature in the X-ray spectra is, indeed, due to absorption, and not intrinsic.

\end{enumerate}


Lastly, the resolution of spectra available from X-ray CCDs is too low to reveal discrete features, such as spectral lines. 
We suspect that only with higher spectral resolution, will one be able to diagnose the X-ray absorbers and determine whether they are intrinsic, part of an (ionized?) outflow, or in the IGM.

\acknowledgments
We acknowledge helpful discussions with Ari Laor and with Eric Feigelson and thank the referee for thoughtful, detailed comments that helped improve the manuscript.
This research is supported by a grant from Israel's Ministry of Science and Technology and by a grant (\#1163/10) from the Israel Science Foundation.


\newpage
\begin{deluxetable}{l c r r c c c c c c r}
\tabletypesize{ \tiny }
\tablecolumns{10} \tablewidth{0pt}
\tablecaption{High-$z$ Quasars}
\tablehead{
   \colhead{Source} &
   \colhead{$z$} &
   \colhead{Photons} &
   \colhead{$N_H$(Gal.)} &
   \colhead{$N_H$($z$)\tablenotemark{a}} &
   \colhead{$\Gamma$\tablenotemark{b}} &
   \colhead{$\log(R)$\tablenotemark{c}} &
   \colhead{$\log(L_R)$\tablenotemark{d}} &
   \colhead{$\log(L_x)$\tablenotemark{e}} &   
   \colhead{$\tau$(0.5 keV)\tablenotemark{f}} &
   \colhead{$\chi ^2$/dof} 
\\
   \colhead{} &
   \colhead{} &
   \colhead{} &
   \colhead{(10$^{20}$cm$^{-2}$)} &
   \colhead{(10$^{22}$cm$^{-2}$)} &
   \colhead{} &
   \colhead{} &
   \colhead{} &
   \colhead{} &
   \colhead{} &
   \colhead{} 
}
\startdata
7C 1428+4218				&	4.715	&	12580	&	1.18	&	$2.1\pm0.5$				&	$1.63\pm0.03$	&	3.3			&	45.0		&	49.3		&		$0.24\pm0.06$	&	1.04		 \\
QSO J0525-3343				&	4.413	&	28800	&	2.28	&	$1.9\pm0.3$				&	$1.63\pm0.02$	&	3.2			&	44.8		&	49.1		&		$0.26\pm0.04$	&	0.98		 \\
QSO B1026-084				&	4.276	&	6250	&	4.49	&	$1.8\pm0.7$				&	$1.42\pm0.04$	&	3.6			&	45.0		&	49.0		&		$0.27\pm0.10$	&	0.98		 \\
QSO B2000-330				&	3.783	&	3300	&	7.22	&	0.80\tablenotemark{*}	&	$1.65\pm0.06$	&	4.0			&	45.9		&	48.5		&		0.14\tablenotemark{*}		&	0.85		 \\	
QSO B0014+810				&	3.366	&	12500	&	13.60	&	$1.8\pm0.5$				&	$1.45\pm0.03$	&	2.3			&	45.1		&	49.2		&		$0.41\pm0.11$	&	0.98		 \\
RX J122135.6+280613			&	3.305	&	3000	&	2.01	&	$0.62\pm0.62$			&	$1.37\pm0.06$	&	2.3			&	43.7		&	48.1		&		$0.15\pm0.15$	&	1.02		 \\
CGRaBS J2129-1538			&	3.280	&	35200	&	4.92	&	$1.6\pm0.2$				&	$1.40\pm0.02$	&	3.4			&	45.0		&	49.6		&		$0.39\pm0.05$	&	0.98		 \\
QSO J0422-3844				&	3.123	&	1800	&	2.11	&	0.90\tablenotemark{*}	&	$1.96\pm0.10$	&	2.2			&	44.5		&	48.1		&		0.24\tablenotemark{*}		&	1.05		 \\
QSO B0537-286				&	3.104	&	14750	&	2.22	&	$0.41\pm0.22$			&	$1.20\pm0.03$	&	4.3			&	45.7		&	49.2		&		$0.11\pm0.06$	&	0.98		 \\
QSO B1524+517				&	2.883	&	2000	&	1.66	&	$2.18\pm0.56$			&	$1.87\pm0.10$	&	-0.2\tablenotemark{*}	&	42.1\tablenotemark{*}	&	47.6		&	$0.71\pm0.18$	&	0.86		 \\
QSO B0438-43				&	2.852	&	7400	&	1.35	&	$1.8\pm0.3$				&	$1.87\pm0.04$	&	5.5			&	46.4		&	48.7		&		$0.59\pm0.12$	&	1.08		 \\
RBS 315						&	2.690	&	69950	&	9.26	&	$2.9\pm0.2$				&	$1.23\pm0.01$	&	3.5			&	44.7		&	49.5		&		$1.08\pm0.06$	&	1.00		 \\
QSO J2354-1513				&	2.675	&	9000	&	2.51	&	$0.6\pm0.2$				&	$1.61\pm0.04$	&	3.9			&	45.4		&	48.1		&		$0.21\pm0.07$	&	1.05		 \\
QSO B1442+2931				&	2.669	&	2450	&	1.41	&	0.53\tablenotemark{*}	&	$1.91\pm0.09$	&	-0.2\tablenotemark{*}	&	42.0\tablenotemark{*}	&	47.5		&		0.18\tablenotemark{*}		&	0.99		 \\
QSO J2220-2803				&	2.406	&	2400	&	1.24	&	0.09\tablenotemark{*}	&	$2.07\pm0.06$	&	-			&	41.9\tablenotemark{*}	&	47.3		&		0.04\tablenotemark{*}		&	0.92		 \\
QSO J0555+3948				&	2.363	&	5000	&	28.20	&	$0.49\pm0.44$			&	$1.52\pm0.05$	&	-			&	46.1		&	48.2		&		$0.22\pm0.21$	&	0.97		 \\
QSO B2149-306				&	2.345	&	36200	&	1.61	&	$0.08\pm0.07$			&	$1.46\pm0.02$	&	3.8			&	45.0		&	48.8		&		$0.036\pm0.032$	&	1.00		 \\
QSO B1318-113				&	2.308	&	2850	&	2.29	&	0.20\tablenotemark{*}	&	$1.98\pm0.07$	&	-			&	41.9\tablenotemark{*}	&	47.0		&		0.09\tablenotemark{*}		&	1.01		 \\
QSO B0237-2322				&	2.225	&	12550	&	2.16	&	$0.10\pm0.10$			&	$1.71\pm0.03$	&	3.7			&	45.8		&	48.2		&		$0.05\pm0.05$	&	1.06		 \\
4C 71.07					&	2.172	&	225450	&	2.85	&	$0.09\pm0.03$			&	$1.33\pm0.01$	&	3.3			&	45.8		&	49.5		&		$0.05\pm0.02$	&	1.08		 \\
QSO J1250+2631				&	2.043	&	4550	&	8.55	&	0.03 \tablenotemark{*}	&	$2.11\pm0.04$	&	-0.4		&	42.1		&	47.3		&		0.02 \tablenotemark{*}		&	1.03		 \\
SDSS J110449.13+381811.6	&	1.942	&	38800	&	1.91	&	$0.13\pm0.05$			&	$2.55\pm0.03$	&	0.4			&	41.7		&	47.3		&		$0.08\pm0.03$	&	1.08		 \\
QSO B2202-0209				&	1.770	&	24050	&	5.50	&	$0.46\pm0.06$			&	$1.78\pm0.02$	&	-			&	41.6\tablenotemark{*}	&	47.9		&		$0.35\pm0.05$	&	1.02		 \\
QSO B1115+080A				&	1.736	&	11800	&	3.73	&	$0.65\pm0.09$			&	$1.86\pm0.04$	&	-0.6\tablenotemark{*}	&	41.6\tablenotemark{*}	&	47.2		&		$0.52\pm0.07$	&	0.88		 \\
QSO J1004+4112A				&	1.734	&	8850	&	1.31	&	0.07\tablenotemark{*}	&	$1.78\pm0.03$	&	0.5\tablenotemark{*}	&	41.6\tablenotemark{*}	&	47.2		&		0.06\tablenotemark{*}		&	0.91		 \\
SDSS J110400.27+380230.9	&	1.622	&	16900	&	1.58	&	0.07\tablenotemark{*}	&	$2.81\pm0.04$	&	0.9			&	41.4		&	47.6		&		0.06\tablenotemark{*}		&	1.01		 \\
RX J111822.1+074450			&	1.618	&	4950	&	3.73	&	0.05\tablenotemark{*}	&	$1.92\pm0.08$	&	0.7\tablenotemark{*}	&	41.5\tablenotemark{*}	&	46.2		&		0.05\tablenotemark{*}		&	0.95		 \\
QSO J1312+2319				&	1.508	&	2250	&	1.13	&	$0.38\pm0.19$			&	$1.67\pm0.09$	&	2.6			&	44.1		&	46.3		&		$0.36\pm0.18$	&	1.07		 \\
XMS J221538.1-174631		&	1.416	&	2650	&	1.94	&	0.06\tablenotemark{*}	&	$2.09\pm0.11$	&	0.1\tablenotemark{*}	&	41.4\tablenotemark{*}	&	45.5		&		0.06\tablenotemark{*}		&	0.92		 \\
QSO B0909+5312				&	1.377	&	8150	&	1.49	&	0.007\tablenotemark{*}	&	$1.77\pm0.03$	&	0.2			&	41.9		&	47.1		&		0.007\tablenotemark{*}		&	0.99		 \\
QSO B1634+7037				&	1.337	&	9850	&	4.97	&	0.03\tablenotemark{*}	&	$2.10\pm0.03$	&	-0.4		&	41.8		&	47.3		&		0.03\tablenotemark{*}		&	1.02		 \\
2XMM J121426.5+140259		&	1.279	&	4250	&	2.71	&	0.04\tablenotemark{*}	&	$1.75\pm0.04$	&	3.6			&	44.1		&	46.3		&		0.05\tablenotemark{*}		&	1.00		 \\
4C 06.41					&	1.270	&	2370	&	2.56	&	0.005\tablenotemark{*}	&	$1.58\pm0.02$	&	2.7			&	44.4		&	47.2		&		0.008\tablenotemark{*}		&	1.01		 \\
SDSS J103031.64+052454.9	&	1.183	&	1950	&	2.53	&	0.04\tablenotemark{*}	&	$2.20\pm0.08$	&	0.1\tablenotemark{*}	&	41.2\tablenotemark{*}	&	45.6		&		0.06\tablenotemark{*}		&	1.03		 \\
LBQS 2212-1747				&	1.159	&	2600	&	2.06	&	0.06\tablenotemark{*}	&	$2.63\pm0.10$	&	0.2\tablenotemark{*}	&	41.1\tablenotemark{*}	&	45.8		&		0.09\tablenotemark{*}		&	0.84		 \\
QSO B0239-0012				&	1.104	&	2450	&	2.93	&	0.03\tablenotemark{*}	&	$1.98\pm0.09$	&	0.4\tablenotemark{*}	&	41.1\tablenotemark{*}	&	45.9		&		0.05\tablenotemark{*}		&	0.78		 \\
SDSS J102313.25+195651.8	&	1.086	&	3700	&	1.98	&	0.05\tablenotemark{*}	&	$1.85\pm0.07$	&	0.4\tablenotemark{*}	&	41.1\tablenotemark{*}	&	45.5		&		0.08\tablenotemark{*}		&	0.97		 \\
QSO B2302+029				&	1.052	&	2000	&	4.98	&	0.02\tablenotemark{*}	&	$2.27\pm0.07$	&	-			&	41.0\tablenotemark{*}	&	46.2		&		0.04\tablenotemark{*}		&	1.00		 \\
SDSS J124938.40+050925.5	&	0.991	&	3850	&	2.02	&	0.02\tablenotemark{*}	&	$2.14\pm0.05$	&	0.6\tablenotemark{*}	&	41.0\tablenotemark{*}	&	45.9		&		0.03\tablenotemark{*}		&	1.16		 \\
XMS J221523.6-174318		&	0.956	&	2050	&	1.94	&	0.03\tablenotemark{*}	&	$2.13\pm0.09$	&	0.1\tablenotemark{*}	&	40.9\tablenotemark{*}	&	45.0		&		0.05\tablenotemark{*}		&	1.04		 \\
QSO B0235+1624				&	0.940	&	44450	&	7.70	&	$0.60\pm0.03$			&	$2.33\pm0.02$	&	2.7			&	44.6		&	47.3		&		$0.39\pm0.02$	&	1.01		 \\
SDSS J134834.28+262205.9	&	0.918	&	9800	&	1.20	&	0.01\tablenotemark{*}	&	$3.12\pm0.04$	&	0.6			&	41.0		&	46.0		&		0.01\tablenotemark{*}		&	1.15		 \\
LBQS 1212+1411				&	0.848	&	1600	&	2.71	&	0.03\tablenotemark{*}	&	$2.36\pm0.10$	&	-1.3\tablenotemark{*}	&	40.8\tablenotemark{*}	&	45.7		&		0.06\tablenotemark{*}		&	0.94		 \\
4C 67.14					&	0.844	&	3900	&	4.28	&	0.02\tablenotemark{*}	&	$1.76\pm0.04$	&	-			&	44.1		&	46.3		&		0.05\tablenotemark{*}		&	0.94		 \\
RX J111750.5+075712			&	0.698	&	3950	&	3.46	&	$0.17\pm0.06$			&	$1.76\pm0.07$	&	0.9\tablenotemark{*}	&	40.6\tablenotemark{*}	&	45.6		&		$0.41\pm0.14$	&	1.06		 \\
2MASS J08404758+1312238		&	0.680	&	3700	&	4.00	&	$0.21\pm0.05$			&	$1.75\pm0.06$	&	3.8			&	44.1		&	46.3		&		$0.53\pm0.14$	&	1.03		 \\
SDSS J144404.50+291412.2	&	0.660	&	2600	&	1.40	&	0.01\tablenotemark{*}	&	$2.85\pm0.08$	&	-0.1\tablenotemark{*}	&	40.5\tablenotemark{*}	&	45.5		&		0.01\tablenotemark{*}		&	1.03		 \\
QSO B0121+318				&	0.654	&	3600	&	5.25	&	0.01\tablenotemark{*}	&	$1.99\pm0.05$	&	2.2			&	42.9		&	45.9		&		0.04\tablenotemark{*}		&	0.99		 \\
SDSS J121952.31+472058.5	&	0.653	&	2550	&	1.50	&	0.01\tablenotemark{*}	&	$1.77\pm0.06$	&	0.6			&	40.5		&	45.3		&		0.03\tablenotemark{*}		&	1.01		 \\
QSO B1218.7+7522			&	0.645	&	3900	&	2.61	&	0.05\tablenotemark{*}	&	$1.66\pm0.09$	&	0.2\tablenotemark{*}	&	40.5\tablenotemark{*}	&	45.6		&		0.13\tablenotemark{*}		&	0.85		 \\
4C 39.27					&	0.618	&	6600	&	1.32	&	0.01\tablenotemark{*}	&	$1.93\pm0.03$	&	2.8			&	43.3		&	45.6		&		0.01\tablenotemark{*}		&	0.98		 \\
QSO B1207+39				&	0.615	&	119700	&	2.23	&	$0.10\pm0.01$			&	$2.23\pm0.01$	&	2.4			&	41.7		&	46.7		&		$0.28\pm0.02$	&	1.06		 \\
XBS J124903.6-061049		&	0.610	&	2900	&	1.99	&	0.08\tablenotemark{*}	&	$2.20\pm0.13$	&	1.7			&	41.6		&	45.0		&		0.22\tablenotemark{*}		&	0.99		 \\
RX J1348.8+2622				&	0.598	&	3000	&	1.22	&	0.01\tablenotemark{*}	&	$3.00\pm0.17$	&	0.3\tablenotemark{*}	&	40.4\tablenotemark{*}	&	45.1		&		0.02\tablenotemark{*}		&	1.60		 \\
SDSS J122532.29+332533.5	&	0.586	&	3100	&	1.82	&	0.05\tablenotemark{*}	&	$1.77\pm0.08$	&	-			&	40.5		&	44.9		&		0.15\tablenotemark{*}		&	0.89		 \\
SDSS J104426.02+063304.5 	&	0.562	&	3300	&	2.53	&	0.01\tablenotemark{*}	&	$2.05\pm0.05$	&	0.4\tablenotemark{*}	&	40.4\tablenotemark{*}	&	45.2		&		0.04\tablenotemark{*}		&	1.02		 \\
SDSS J111135.76+482945.3	&	0.558	&	4350	&	1.72	&	0.01\tablenotemark{*}	&	$2.07\pm0.05$	&	0.2\tablenotemark{*}	&	40.4\tablenotemark{*}	&	45.5		&		0.04\tablenotemark{*}		&	1.02		 \\
QSO B1334-127				&	0.539	&	14400	&	5.16	&	$0.03\pm0.02$			&	$1.75\pm0.03$	&	3.5			&	44.4		&	46.3		&		$0.10\pm0.06$	&	0.96		 \\
2MASS J10461372+5255544		&	0.503	&	4450	&	1.16	&	0.02\tablenotemark{*}	&	$3.67\pm0.08$	&	-0.1\tablenotemark{*}	&	40.2\tablenotemark{*}	&	45.8		&		0.06\tablenotemark{*}		&	0.86		 \\
RX J1626.4+3513				&	0.497	&	3700	&	1.40	&	$0.05\pm0.03$			&	$2.72\pm0.10$	&	1.8			&	41.6		&	45.7		&		$0.16\pm0.11$	&	0.94		 \\
QSO B1157-1942				&	0.450	&	36050	&	3.13	&	0.001\tablenotemark{*}	&	$1.76\pm0.01$	&	0.8			&	41.1		&	46.1		&		0.006\tablenotemark{*}		&	1.00		 \\

\tablenotetext{a}{~Measured assuming an absorber at the host at redshift $z$.}
\tablenotetext{b}{~Photon index of power law.}
\tablenotetext{c}{~Radio loudness, calculated using the definition $R \equiv \nu L_{\nu} (5 \mathrm{GHz})/\nu L_{\nu} (4400 \mathrm{\AA})$.}
\tablenotetext{d}{~Radio luminosity $\nu L_{\nu}$ in $\rm erg~s^{-1}$ at 5~GHz calculated using $L_R=4\pi d^2 F_R$, where $d$ is the luminosity distance.}
\tablenotetext{e}{~Unabsorbed X-ray luminosity in $\rm erg~s^{-1}$ (0.2 to 20 keV) calculated using $L_x=4\pi d^2 F_x$, where $d$ is the luminosity distance.}
\tablenotetext{f}{~Deduced from $N_H (z)$ using Eq.~\ref{eq:tau_equation}.}
\tablenotetext{*}{~Upper limit.}

\enddata

\label{table_of_objects}
\end{deluxetable}

\newpage
\begin{table}
\centering
\caption[Summary of correlation tests]{\centering Summary of correlation tests}
\setlength{\tabcolsep}{0.7ex} 
{\scriptsize
\begin{tabular}{| l | c | c | c | c |}

\hline

\multirow{2}{*}				&	\multicolumn{2}{|c|}{Generalized Kendall's $\tau$} 			&	\multicolumn{2}{|c|}{Generalized Spearman's $\rho$} 			\\
\cline{2-3} \cline{4-5}

							&	$Z$ value					&	Chance probability			&	$\rho$							&	Chance probability			\\	
\hline

$\log(\tau)~vs.~\log(1+z)$	&	3.150						&	0.0016						&	0.389							&	0.0033						\\
$\log(\tau)~vs.~\log(R)$	&	2.989						&	0.0028						&	0.423							&	0.0028						\\
$\log(\tau)~vs.~\log(L_R)$	&	2.861						&	0.0042						&	0.423							&	0.0014						\\

\hline

\end{tabular}
}

\label{table_of_correlation_tests}
\end{table}

\newpage
\begin{deluxetable}{l c r c r c c}
\tabletypesize{ \scriptsize }
\tablecolumns{6} \tablewidth{0pt}
\tablecaption{Broken power-law fit for absorbed objects}
\tablehead{
   \colhead{Source} &
   \colhead{$z$} &
   \colhead{$N_H$(Gal.)} &
   \colhead{Observed break energy} &
   \colhead{$\Delta \Gamma$} &
   \colhead{$\chi ^2$/dof \tablenotemark{a}} &
   \colhead{$\chi ^2$/dof \tablenotemark{b}} 
\\
   \colhead{} &
   \colhead{} &
   \colhead{(10$^{20}$cm$^{-2}$)} &
   \colhead{(keV)} &
   \colhead{($\Gamma_\mathrm{high}-\Gamma_\mathrm{low}$)} &
   \colhead{(Broken)} &
   \colhead{(Absorbed)} 
}
\startdata
7C 1428+4218				&	4.715	&			1.18				&	$0.91_{-0.24}^{+0.29}$		&	$0.36 \pm 0.21$		&	0.95		 		&	1.04				\\
QSO J0525-3343				&	4.413	&			2.28				&	$0.51_{-0.05}^{+0.31}$		&	$1.78 \pm 0.21$		&	0.98			 	&	0.98				\\
QSO B1026-084				&	4.276	&			4.49				&	$0.85_{-0.30}^{+0.46}$		&	$0.46 \pm 0.82$		&	0.90			 	&	0.98				\\
QSO B0014+810				&	3.366	&			13.60				&	$0.77_{-0.14}^{+0.26}$		&	$0.73 \pm 0.46$		&	0.93			 	&	0.98				\\
RX J122135.6+280613			&	3.305	&			2.01				&	$0.63_{-0.13}^{+0.47}$		&	$0.86 \pm 1.36$		&	1.15			 	&	1.02				\\
CGRaBS J2129-1538			&	3.280	&			4.92				&	$0.83_{-0.16}^{+0.11}$		&	$0.59 \pm 0.28$		&	0.99			 	&	0.98				\\
QSO B0537-286				&	3.104	&			2.22				&	$0.55_{-0.04}^{+0.35}$		&	$0.64 \pm 0.83$		&	0.98			 	&	0.98				\\
QSO B1524+517				&	2.883	&			1.66				&	$0.98_{-0.11}^{+0.19}$		&	$1.04 \pm 0.27$		&	0.80		 		&	0.86				\\
RBS 315						&	2.690	&			9.26				&	$0.95_{-0.04}^{+0.06}$		&	$1.18 \pm 0.12$		&	1.02			 	&	1.00				\\
QSO J2354-1513				&	2.675	&			2.51				&	$0.50_{-0.03}^{+0.21}$		&	$1.78 \pm 0.62$		&	1.04			 	&	1.05				\\
QSO J0555+3948				&	2.363	&			28.20				&	$2.59_{-0.64}^{+2.12}$		&	$-0.39 \pm 0.27$	&	0.99			 	&	0.97				\\
QSO B2149-306				&	2.345	&			1.61				&	$3.11_{-0.99}^{+5.46}$		&	$-0.10 \pm 0.06$	&	0.99			 	&	1.00				\\
4C 71.07					&	2.172	&			2.85				&	$9.48_{-0.42}^{+0.60}$		&	$-1.17 \pm 0.66$	&	1.06			 	&	1.08				\\
SDSS J110449.13+381811.6	&	1.942	&			1.91				&	$0.72_{-0.07}^{+0.13}$		&	$0.36 \pm 0.11$		&	1.12			 	&	1.08				\\
QSO B2202-0209				&	1.770	&			5.50				&	$0.59_{-0.06}^{+0.04}$		&	$1.34 \pm 0.56$		&	1.11			 	&	1.02				\\
QSO B1115+080A				&	1.736	&			3.73				&	$1.00_{-0.12}^{+0.12}$		&	$0.73 \pm 0.14$		&	0.84			 	&	0.88				\\
QSO B0235+1624				&	0.940	&			7.70				&	$0.72_{-0.02}^{+0.02}$		&	$2.31 \pm 0.19$		&	1.13		 		&	1.01				\\
RX J111750.5+075712			&	0.698	&			3.46				&	$0.73_{-0.10}^{+0.14}$		&	$1.19 \pm 0.34$		&	1.30			 	&	1.06				\\
2MASS J08404758+1312238		&	0.680	&			4.00				&	$0.77_{-0.09}^{+0.10}$		&	$1.07 \pm 0.37$		&	1.09			 	&	1.03				\\
QSO B1207+39				&	0.615	&			2.23				&	$1.00_{-0.05}^{+0.07}$		&	$0.44 \pm 0.03$		&	1.02			 	&	1.06				\\
QSO B1334-127				&	0.539	&			5.16				&	$0.79_{-0.26}^{+0.36}$		&	$0.17 \pm 0.20$		&	0.96		 		&	0.96				\\
RX J1626.4+3513				&	0.497	&			1.40				&	$1.52_{-0.46}^{+0.44}$		&	$0.46 \pm 0.27$		&	0.90			 	&	0.94				\\

\tablenotetext{a}{~Goodness of fit for the broken power-law model.}
\tablenotetext{b}{~Goodness of fit for the absorbed power-law model.}

\enddata
\label{table_of_break_energy}
\end{deluxetable}

%
%

\newpage

\newpage

\newpage

\newpage

\newpage

\newpage

%
%

\newpage

\newpage

%
%

\newpage

%
%

\newpage

\newpage

\newpage


\end{document}